\documentclass{aa}
\usepackage[varg]{txfonts}
\usepackage{graphicx,grffile, booktabs}
\usepackage{amssymb,float,natbib}
\usepackage[draft]{hyperref}
\bibpunct{(}{)}{;}{a}{}{,} 

\newcommand{\kms}{km s$^{-1}$}
\newcommand{\ceo}{C$^{18}$O}
\newcommand{\tco}{$^{13}$CO}

\newcommand{\mco}{$^{12}$CO}

\begin{document}

\title{Rotationally-supported disks around Class I sources in Taurus: disk 
formation constraints
\thanks{Based on observations carried out with the IRAM Plateau de Bure 
Interferometer. IRAM is supported by INSU/CNBRS(France), MPG
   (Germany) and IGN (Spain).}}
\author{D.~Harsono \inst{\ref{inst1}, \ref{inst2} } \and 
J.~K.~J{\o}rgensen \inst{\ref{inst3}, \ref{inst4}} \and E.~F.~van Dishoeck 
\inst{\ref{inst1}, \ref{inst5}} 
\and M.~R.~Hogerheijde \inst{\ref{inst1}} \and 
S.~Bruderer \inst{\ref{inst5}} \and 
M.~V.~Persson \inst{\ref{inst1}, \ref{inst3}, \ref{inst4}} 
\and J.~C.~Mottram \inst{\ref{inst1}}
}

\institute{Leiden Observatory, Leiden University, P.O. Box 9513, 2300 RA Leiden, 
The Netherlands {\email: harsono@strw.leidenuniv.nl} \label{inst1} 
\and
SRON Netherlands Institute for Space Research, PO Box 800, 9700 AV, Groningen,
The Netherlands \label{inst2}
\and
Niels Bohr Institute, University of Copenhagen, Juliane Maries Vej 30, 2100  
Copenhagen {\O}, Denmark \label{inst3}
\and 
Centre for Star and Planet Formation, Natural History Museum of Denmark,  
University of Copenhagen, {\O}ster Voldgade 5-7, 1350 Copenhagen K, Denmark 
\label{inst4}
\and
Max-Planck-Institut f{\" u}r extraterretrische Physik, Giessenbachstrasse 1,
85748, Garching, Germany \label{inst5}
}

\abstract{Disks are observed around pre-main sequence stars, but how and 
when they form is still heavily debated.  While disks around young stellar 
objects have been identified through thermal dust emission, spatially and 
spectrally resolved molecular line observations are needed to determine their 
nature.  Only a handful of embedded rotationally supported disks have been  
identified to date. }
{We identify and characterize rotationally supported disks near the end of the 
main accretion phase of low-mass protostars by comparing their gas and dust 
structures.}  
{Subarcsecond observations of dust and gas toward four Class I low-mass young 
stellar objects in Taurus are presented at significantly higher sensitivity than 
previous studies.  The \tco\ and \ceo\ $J$ = 2--1 transitions at 220 GHz were 
observed with the Plateau de Bure Interferometer at a spatial resolution of 
$\le$0.8$\arcsec$ (56 AU radius at 140 pc) and analyzed using $uv$-space 
position velocity diagrams to determine the nature of their observed velocity 
gradient.}
{Rotationally supported disks (RSDs) are detected around 3 of the 4 Class 
I sources studied.  The derived masses identify them as Stage I objects; i.e., their 
stellar mass is higher than their envelope and disk masses. The outer radii of 
the Keplerian disks toward our sample of Class I sources are $\leq 100$ AU.  The 
lack of on-source \ceo\ emission for TMR1 puts an upper limit of 50 AU on its 
size.  Flattened structures at radii $>100$ AU around these sources are 
dominated by infalling motion ($\upsilon \propto r^{-1}$).  A large-scale 
envelope model is required to estimate the basic parameters of the 
flattened structure from spatially resolved continuum data.  Similarities and 
differences between the gas and dust disk are discussed. 
Combined with literature data, the sizes of the RSDs around Class I objects are 
best described with evolutionary models with an initial rotation of $\Omega = 
10^{-14}$ Hz and slow sound speeds.  Based on the comparison of gas and dust 
disk masses, little CO is frozen out within 100 AU in these disks. }
{Rotationally supported disks with radii up to 100 AU are present around Class 
I embedded objects. Larger surveys of both Class 0 and I objects are needed to 
determine whether most disks form late or early in the  embedded phase.  }


\titlerunning{Velocity structure of Class I disks}
\authorrunning{D. Harsono et al.}

\maketitle


\section{Introduction}\label{sec:intro}

Rotationally supported accretion disks (RSDs) are thought to form very early 
during the star formation process \citep[see, e.g.,][]{bodenheimer95}.  The RSD 
transports a significant fraction of the mass from the envelope onto the 
young star and eventually evolves into a protoplanetary disk as the envelope 
dissipates.   The presence of an RSD affects not only the physical structure 
of the system but also the chemical content and evolution as it promotes the 
production of more complex chemical species by providing a longer lifespan at 
mildly elevated temperature \citep{aikawa08, visser11, aikawa11}.  Although the 
standard picture of star formation predicts early formation and evolution of 
RSDs, theoretical studies suggest that the presence of magnetic fields 
prevents the formation of RSDs in the early stages of star formation 
\citep[e.g.,][]{galli03, chiang08, zyli11, joos12}.  Unstable flattened disk-like 
structures are formed in such simulations, and indeed, disks in the embedded 
phase have been inferred through continuum observations \citep[e.g.,][]
{looney03, prosac09, enoch11, chiang12}.  However, with only continuum data, it 
is difficult to distinguish between such a feature and an RSD.  Thus, spatially and 
spectrally resolved observations of the gas are needed to unravel the nature of 
these embedded disks.

Only a handful of studies have explored the change in the velocity profiles 
from the envelope to the disk with spectrally resolved molecular lines 
\citep{lee10, yen13, murillo13b}.  It is essential to differentiate between the disk 
and the infalling envelope through such analysis.  This paper presents IRAM 
Plateau de Bure Interferometer (PdBI) observations of \tco\ and \ceo\ $J=$2--1 
toward four Class I sources in Taurus ($d = 140$ pc) at higher angular resolution 
($\sim$0.8$\arcsec = 112$ AU diameter at 140 pc) and higher sensitivity than 
previously possible.  We aim to study the velocity profile as revealed through 
these molecular lines and constrain the size of embedded RSDs toward these 
sources.

 The embedded phase can be divided into the following stages  
\citep{robitaille06}: Stage 0 ($M_{\rm env} > M_{\star}$), Stage I ($M_{\rm env} 
< M_{\star}$ but $M_{\rm disk} < M_{\rm env}$) and Stage II ($M_{\rm env} < 
M_{\rm  disk}$), where $M_{\star}$, $M_{\rm env}$ and $M_{\rm disk}$ are masses 
of the central protostar, envelop and RSD respectively.  This is not to be 
confused with Classes, which are observationally derived evolutionary 
indicators \citep{lada84, andre93} that do not necessarily trace evolutionary 
stage due to geometrical effects \citep{whitney03a, crapsi08, dunham10}.  
However, since all Class I sources discussed in this paper also turn out to be 
true Stage I sources, we use the Class notation throughout this paper for
convenience.

RSDs are ubiquitous once most of the envelope has been dissipated away (Class 
II) with outer radius $R_{\rm out}$ up to approximately 200 AU \citep[][and 
references therein]{wc11}.  On the other hand, the general kinematical structure 
of deeply embedded protostars on small scales is still not well understood. There 
is growing evidence of rotating flattened structures around Class I sources 
\citep{km90, hayashi93, ohashi97a, ohashi97b, brinch07, lommen08, prosac09, 
lee10, lee11, takakuwa12, yen13}, but the question remains how and when RSDs 
form in the early stages of star formation and what their sizes are.

Class I young stellar objects (YSOs) are ideal targets to search for embedded 
RSDs.  At this stage, the envelope mass has substantially decreased such that the 
embedded RSD dominates the spatially resolved CO emission \citep{harsono13a}.  
CO is the second most abundant molecule and is chemically stable, thus the disk 
emission should be readily detected.  Furthermore, \citet{harsono13a} showed 
that the size of the RSD can be directly measured by analyzing the velocity profile 
observed through spatially and spectrally resolved CO observations.  The sources 
targeted here are TMC1 (\object{IRAS 04381+2540}), TMR1 (\object{IRAS 
04361+2547}), L1536 (\object{IRAS 04295+2251}), and TMC1A (\object{IRAS 
04365+2535}).  These Class I objects have been previously observed by the 
Submillimeter Array (SMA) and the Combined Array for Research in Millimeter-wave 
Astronomy (CARMA) at lower sensitivity and/or resolution 
\citep[e.g.,][]{prosac09, eisner12, yen13}.  Embedded rotating flattened 
structures have been proposed previously around TMC1 and TMC1A \citep{ohashi97a, 
hogerheijde98, brown99, yen13}.  Thus, we target these sources to determine the 
presence of Keplerian disks, constrain their sizes near the end of the main 
accretion phase on scales down to 100 AU and compare this with the dust 
structure.

This paper is laid out as follows.  Section~\ref{sec:obs} presents the 
observations and data reduction.  The continuum and line maps are presented in 
Section~\ref{sec:obsres}.  In Section~\ref{sec:analysis}, disk masses and sizes 
are obtained through continuum visibility modelling.  Position-velocity diagrams 
are then analyzed to determine the size of rotationally supported disks.  
Section~\ref{sec:dis} discusses the implications of the observed rotational 
supported disks toward Class 0 and I sources.  Section~\ref{sec:sum} summarizes 
the conclusions of this paper.


\section{Observations}\label{sec:obs}


\begin{table*}[htb!]
 \caption{Noise levels and beam sizes.}
 \centering
 \label{tbl:weights}
 \begin{tabular}{c c  c c c c c c }\toprule \hline
  & & \multicolumn{2}{c}{Continuum} & & & Lines & \\ \cline{3-4} \cline{6-8}
 Weighting & Beam & RMS$_{\rm th}$\tablefootmark{a} & RMS$_{\rm 
cl}$\tablefootmark{b} & &  RMS$_{\rm th}$ & \tco\ RMS$_{\rm cl}$ & \ceo\ 
RMS$_{\rm cl}$ \\
  & size (PA) & \multicolumn{2}{c}{[mJy beam$^{-1}$]} & & 
\multicolumn{3}{c}{[mJy beam$^{-1}$ channel$^{-1}$]} \\
 \hline
 \multicolumn{8}{c}{TMC1A (TMR1)} \\
 Natural & $1.3\arcsec \times 0.7\arcsec$ (16$^{\circ}$) & 0.12 & 3 (0.6) & & 
24 & 30 (30) & 20 (20) \\
 Uniform & $0.8\arcsec \times 0.7\arcsec$ (177$^{\circ}$) & 0.13 & 3 (0.6) & & 
26 & 30 (29) & 19 (20) \\
 \hline 
 \multicolumn{8}{c}{L1536 (TMC1)} \\
 Natural & $0.7\arcsec \times 0.5\arcsec$ (36$^{\circ}$) & 0.14 & 1 (0.2) & & 
26 & 20 (26) & 18 (18) \\
 Uniform & $0.6\arcsec \times 0.5\arcsec$ (55$^{\circ}$) & 0.13 & 1 (0.2) & & 
29 & 20 (24) & 18 (18) \\
 \bottomrule
 \end{tabular}\\
 \tablefoot{
 \tablefoottext{a}{Theoretical noise (RMS)}.
 \tablefoottext{b}{Peak of cleaned map noise for TMC1A or L1536. TMR1 or 
TMC1 cleaned noise is in parentheses.  The cleaned noise toward TMR1 and TMC1 
are significantly lower due to weaker total continuum flux.}}
\end{table*}


\subsection{IRAM PdBI Observations}

The sources were observed with IRAM PdBI in the two different configurations 
tabulated in Table~\ref{tbl:obsdetails}.   Observations with {\it an on-source} 
time of $\sim 3.5$ hours per source with baselines from 15 to 445 m (11 
k$\lambda$ to 327 k$\lambda$) were obtained toward TMC1A and TMR1 in 
March 2012.  Additional track-sharing observations of L1536 and TMC1 were 
obtained in March-April 2013 with an {\it on-source} time of $\sim 3$ hours per 
source and baselines between 20 to 450 m (15 k$\lambda$ to 331 
k$\lambda$).  The receivers were tuned to 219.98 GHz (1.36 mm) in order to 
simultaneously observe the \tco\ (220.3987 GHz) and \ceo\ (219.5603 GHz) 
$J=$2--1 lines.  The narrow correlators (bandwidth of 40 MHz $\sim 54$ \kms ) 
were centered on each line with a spectral resolution of 0.078 MHz (0.11 \kms).  
In addition, the WideX correlator was used, which covers a 3.6 GHz window 
at a resolution of 1.95 MHz (2.5--3 \kms) with 6 mJy beam$^{-1}$ 
channel$^{-1}$ RMS.

The calibration and imaging were performed using the \textsc{CLIC} and 
{\scshape MAPPING} packages of the IRAM {\scshape GILDAS} 
software\footnote{http://www.iram.fr/IRAMFR/GILDAS}.  The standard calibration 
was followed using the calibrators tabulated in Table~\ref{tbl:obsdetails}.  The 
data quality assessment tool flagged out any integrations with significantly 
deviating amplitude and/or phase and the continuum was subtracted from the line 
data before imaging.  The continuum visibilities were constructed using the 
WideX correlator centered on 219.98 GHz (1.36 mm).  The resulting beam sizes and 
noise levels for natural and uniform weightings can be found in 
Table~\ref{tbl:weights}.  The uniform weighted images will be used for the 
analysis in the image space.


\subsection{WideX detections}

The WideX correlators cover frequencies between 218.68 -- 222.27 GHz.  Strong 
molecular lines that are detected toward all of the sources within the WideX 
correlator include \ceo\ 2--1 (219.5603541 GHz), SO $5_6$--$4_5$ (219.9488184 
GHz), and \tco\ 2--1 (220.3986841 GHz).  The WideX spectra and integrated flux 
maps are presented in Appendix~\ref{app:widex} but not analyzed here.


\section{Results}\label{sec:obsres}

\subsection{Continuum maps}


\setlength{\tabcolsep}{3.5pt}
\begin{table*}[htb!]
 \centering
 \caption{1.36 mm continuum properties from elliptical Gaussian fits and 
derived disk and envelope masses (within 15$''$) from fluxes.}
 \label{tbl:sources}
 \begin{tabular}{lccccccccccc}\toprule \hline
  Source & RA & Dec & $\upsilon_{\rm lsr}$ & $T_{\rm bol}$\tablefootmark{a} & 
$L_{\rm bol}$\tablefootmark{a} & Size (PA)\tablefootmark{b} & $F_{\rm 1.36 mm}$ 
& $S_{\rm 1.3 mm}^{\rm int}$\tablefootmark{c} & $S_{\rm 850 \mu m}$ 
(15$\arcsec$) & $M_{\rm env}$ & $M_{\rm disk}$ 
\\
   & (J2000) & (J2000) & [\kms] & [K] & $L_{\odot}$ & [$\arcsec$] ($^{\circ}$) 
& [mJy] & [mJy] & [mJy bm$^{-1}$] & [$M_{\odot}$] & [$M_{\odot}$] \\
   \hline

  TMC1A & 04 39 35.20 & +25 41 44.27 & +6.6 & 118 & 2.7 & $0.45\times0.25$ 
($80$) & $164$ & 450 & 780\tablefootmark{e} & 0.12 & 0.033 \\

  TMC1 & 04 41 12.70 & +25 46 34.80 & +5.2 & 101 & 0.9 & $0.8\times 0.3$ ($84$) 
&   $22$ & 300 & 380\tablefootmark{d} & 0.14 & 0.0039 \\ 
  
  TMR1 & 04 39 13.91 & +25 53 20.57 & +6.3 & 133 & 3.8 & $0.2$ & $50$ &  440 & 
480\tablefootmark{e} & 0.10 & 0.011 \\

  L1536\tablefootmark{f} & 04 32 32.07 & +22 57 26.25 & +5.3 & 270 & 0.4 & $1.1 
\times 0.6$ ($69$) &   $83$ & 115 & 200\tablefootmark{g} & ... & 0.021 \\ 
 \bottomrule
 \end{tabular}\\
 \tablefoot{
 \tablefoottext{a}{Values from \citet{kristensen12}.}
 \tablefoottext{b}{Gaussian fit is used for TMR1.} 
 \tablefoottext{c}{Integrated 1.3 mm flux from \citet{motte01}.}
 \tablefoottext{d}{Peak flux of JCMT SCUBA maps from \citet{francesco08}.}
 \tablefoottext{e}{Peak flux from \citet{prosac09}.}
 \tablefoottext{f}{Position of L1536(W) is listed. }   
 \tablefoottext{g}{Peak flux from \citet{young03}.}
 }
\end{table*}


\begin{figure}[!tbh]
 \centering
 \includegraphics{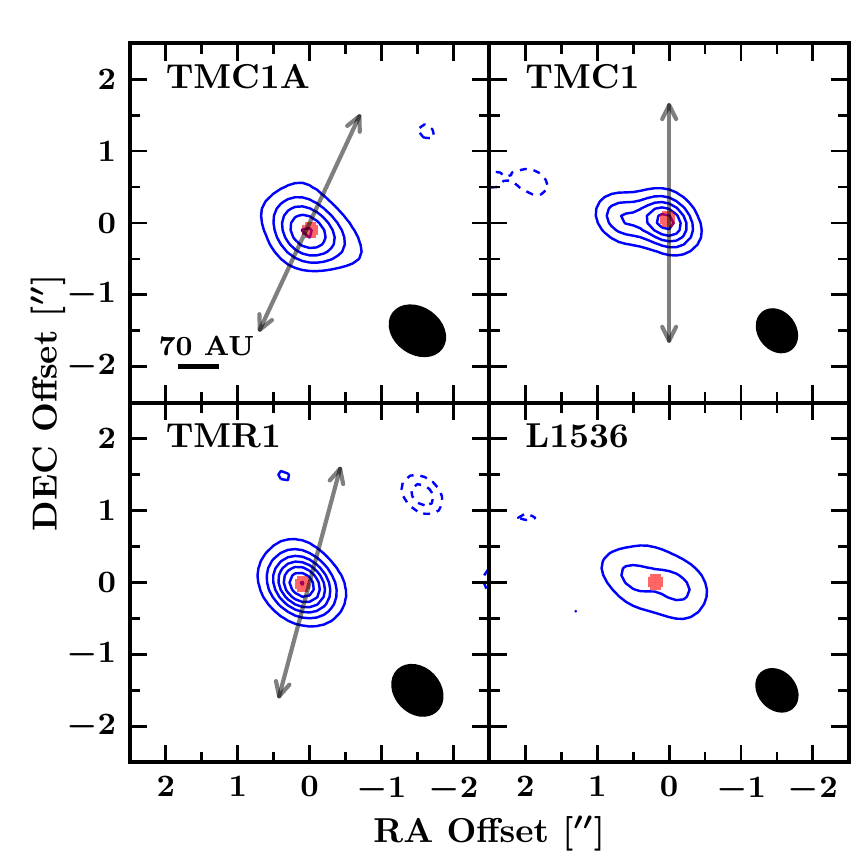}
\caption{1.36 mm uniform weighted continuum maps toward TMC1A 
(I4365+2535, top left),  TMC1 (I4381+2540, top right), TMR1 
(I4361+2547, bottom left), and L1536 (I4295+2251, bottom right).   The 
contours start from 10$\sigma$ up to 
80$\sigma$ by 10$\sigma$.  The peak flux densities per beam are 11 mJy, 42 mJy, 
23 mJy and 128 mJy for TMC1, TMR1, L1536 and TMC1A, respectively.  The red 
circles indicate the best-fit source-position assuming elliptical Gaussian 
while the black ellipses show the synthesized beams.  The arrows indicate the 
outflow direction.  The RMS is given in Table~\ref{tbl:weights}.  The dashed 
blue lines indicate the negative contours starting at 3$\sigma$.}
\label{fig:contmaps}
\end{figure}

\begin{figure}[!tbh]
 \centering
 \includegraphics{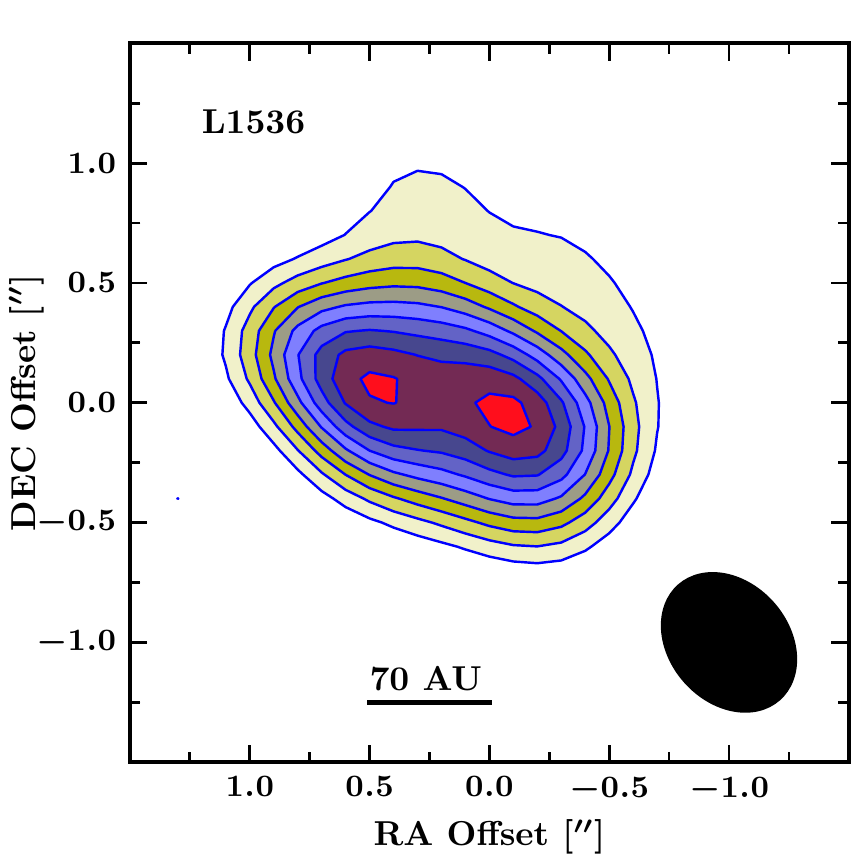}
\caption{Zoom of the continuum image toward L1536.  Both of the line and color 
contours are drawn at 10\% of the peak starting from 4$\sigma$ up to the peak 
intensity of 23 mJy beam$^{-1}$, while the synthesized beam is indicated 
by the black ellipse in the bottom right corner. }
\label{fig:l1536zoom}
\end{figure}


Strong 1.36 mm continuum emission is detected toward all four sources up to 
$\sim$450 m baselines ($\sim 330$ k$\lambda$, which corresponds to roughly 
80 AU in diameter at the distance of Taurus).  To estimate the continuum flux 
and the size of the emitting region, we performed elliptical Gaussian fits to the 
visibilities ($>25$ k$\lambda$).  The best fit parameters are listed in 
Table~\ref{tbl:sources}.  Our derived PdBI 1.36 mm fluxes are $<30$\%  
lower than the single-dish fluxes tabulated in \citet{motte01}.  The only 
exception is L1536 for which 72\% of the singe-dish flux is recovered, which 
indicates a lack of resolved out large-scale envelope around L1536.  The 
deconvolved sizes indicate a large fraction of the emission is within $< 100$ AU 
diameter, consistent with compact flattened dust structures.

Figure~\ref{fig:contmaps} presents the uniform weighted continuum maps with red 
circles indicating the continuum positions (Table~\ref{tbl:sources}).  The 
total flux of TMC1A is $\sim 30$\% lower than in \citet{yen13} within a 
4.0$\arcsec$ $\times$ 3.5$\arcsec$ beam, which indicates that some extended 
structure is resolved-out in our PdBI observations.  The peak fluxes of our 
cleaned maps agree to within 15\% with those reported by \citet{eisner12} except 
toward L1536, which is a factor of two lower in our maps.  However, our image 
toward L1536 (Fig.~\ref{fig:l1536zoom}) shows that there are two peaks whose sum 
is within 15\% of the peak flux reported in \citet{eisner12}.  The position in 
Table~\ref{tbl:sources} is centered on the Western peak and the two peaks are 
separated by $\sim 70$ AU.  The analysis in the next section 
(Section~\ref{sec:analysis}) is with respect to the Western peak.  The maps 
indicate the presence of elongated flattened structure perpendicular to the 
outflow direction found in the literature \citep[TMC1: 0$^{\circ}$; TMR1: 
165$^{\circ}$; TMC1A: 155$^{\circ}$; L1536: None\footnote{Recent \mco\ 3--2 
observations with the JCMT do not show any indication of a bipolar outflow 
from this source, see Appendix~\ref{app:l1536}.},][]{ohashi97a, hogerheijde98, 
brown99}.  Such elongated dust emission is indicative of flattened 
disk-like structures.


\subsection{Molecular lines: maps and morphologies}\label{sec:molmaps}

\begin{figure*}[!tbh]
 \centering
 \begin{tabular}{cc}
 \includegraphics{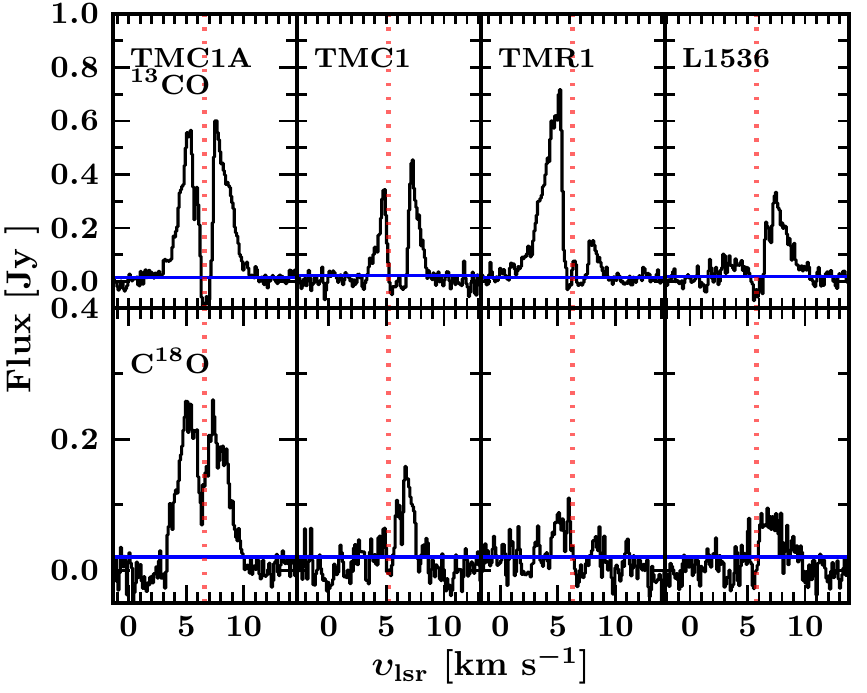}
&
\includegraphics{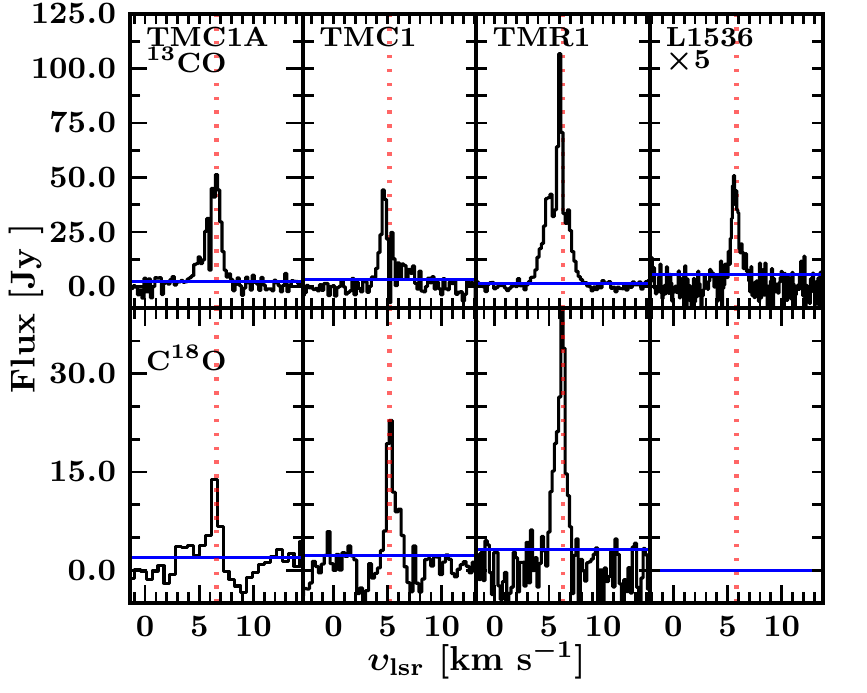}
 \end{tabular}
 \caption{Left: \tco\ (top) and \ceo\ 2--1 (bottom) spectra integrated within a 
1$\arcsec$ box around the continuum position.  Right: Single-dish spectra of 
\tco\ 3--2 (except for L1536, which is \tco\ 2--1) and \ceo\ 2--1.  The vertical 
red dotted lines indicate the systemic velocities of each source derived from 
single dish \ceo\ and C$^{17}$O data \citep{yildiz13a}.  The horizontal blue 
lines indicate the spectral $\sigma_{\rm rms}$.}
\label{fig:lines}
\end{figure*}

\begin{table}[htb!]
 \centering
 \caption{\tco\ and \ceo\ 2--1 integrated flux densities [Jy km s$^{-1}$] 
within a 5$\arcsec$ box around continuum position.}
 \label{tbl:flux}
 \begin{tabular}{c c c c c}\toprule \hline
  Line & \multicolumn{4}{c}{Sources} \\
   & TMC1A & TMC1 & TMR1 & L1536  \\
  \hline
  \tco\ 2--1\tablefootmark{a} & $12.2 \pm 0.08$  & $4.5 \pm 0.07$ & $10.5 \pm 
0.06$ & $3.2 \pm 0.08$  
  \\
  \ceo\ 2--1\tablefootmark{a} &  $5.3 \pm 0.08$  & $2.2 \pm 0.09$ & $3.0 \pm 
0.06$ &  $1.2 \pm 0.08$  \\
\bottomrule
\end{tabular}
\tablefoot{
\tablefoottext{a}{Integrated flux error is calculated through 
1.2$\times\sigma_{\rm 
rms} \sqrt{FWZI \times \delta \upsilon}$.  $FWZI$ is the full width zero 
intensity as calculated from the number of channels $> 5\sigma$.  }
}
\end{table}

\begin{figure*}[!tbh]
 \centering
 \begin{tabular}{cc}
 \includegraphics{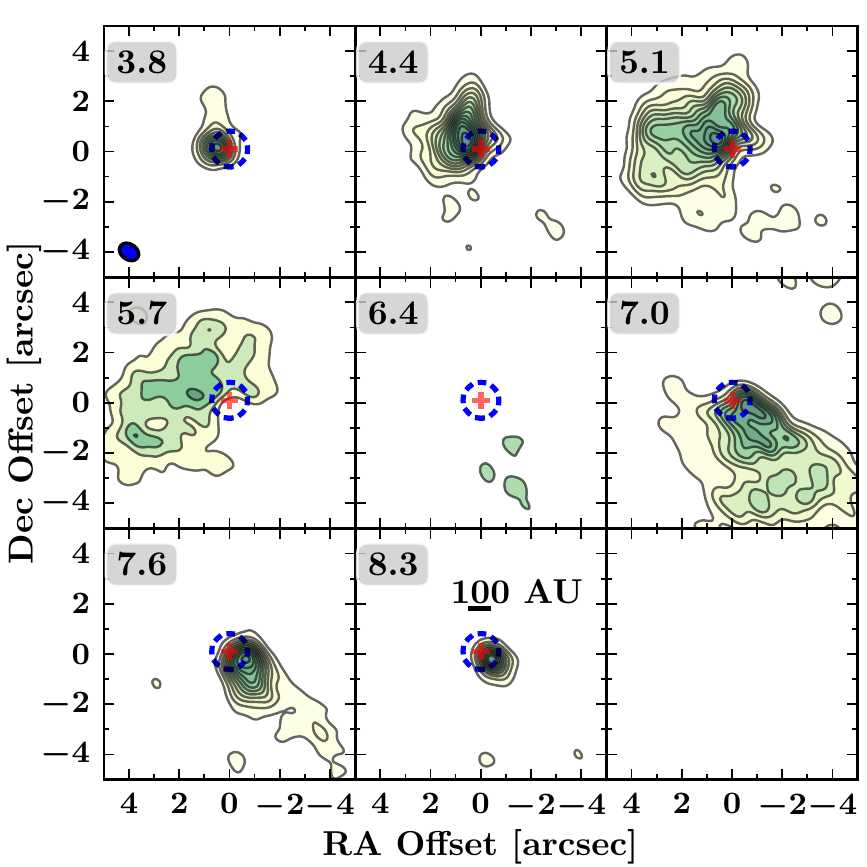}
 &
 \includegraphics{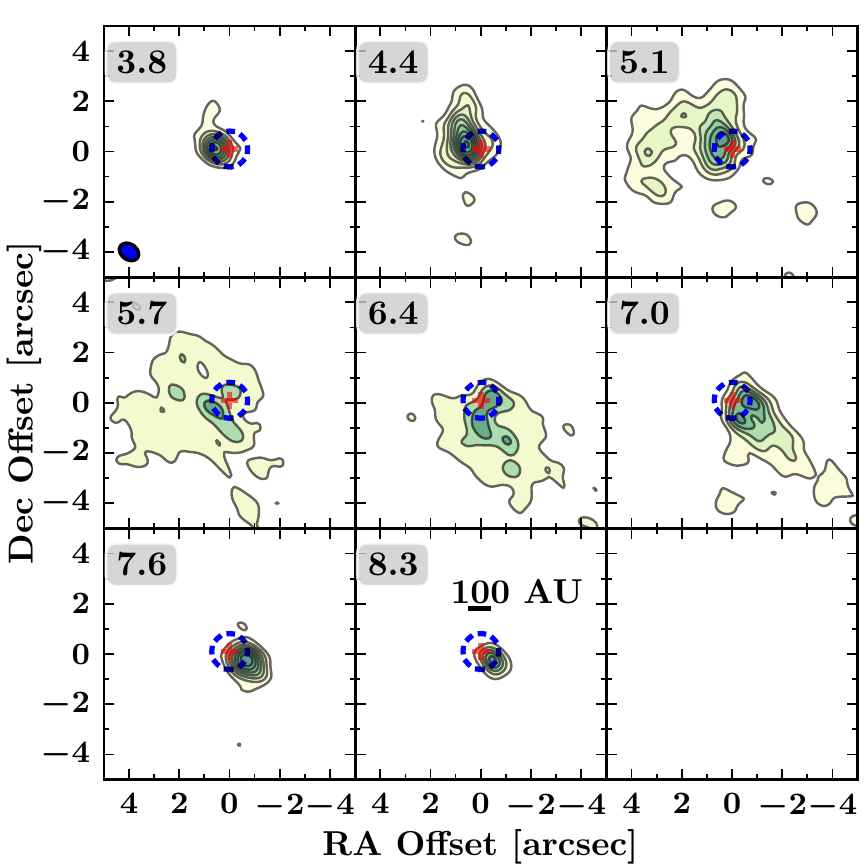}
 \end{tabular}
\caption{ \tco\ (left) and \ceo\ (right) 2--1 channel maps within the inner 
10$\arcsec$ toward TMC1A.  The velocities in \kms\ are indicated at the top 
left in each panel.  The contours are drawn at every 3$\sigma$ starting from 
5$\sigma$ where $\sigma = 12$ mJy km s$^{-1}$ for \tco\ and 10 mJy km s$^{-1}$ 
for \ceo. }
\label{fig:onemap}
\end{figure*}

\begin{figure*}[!tbh]
 \centering
 \includegraphics{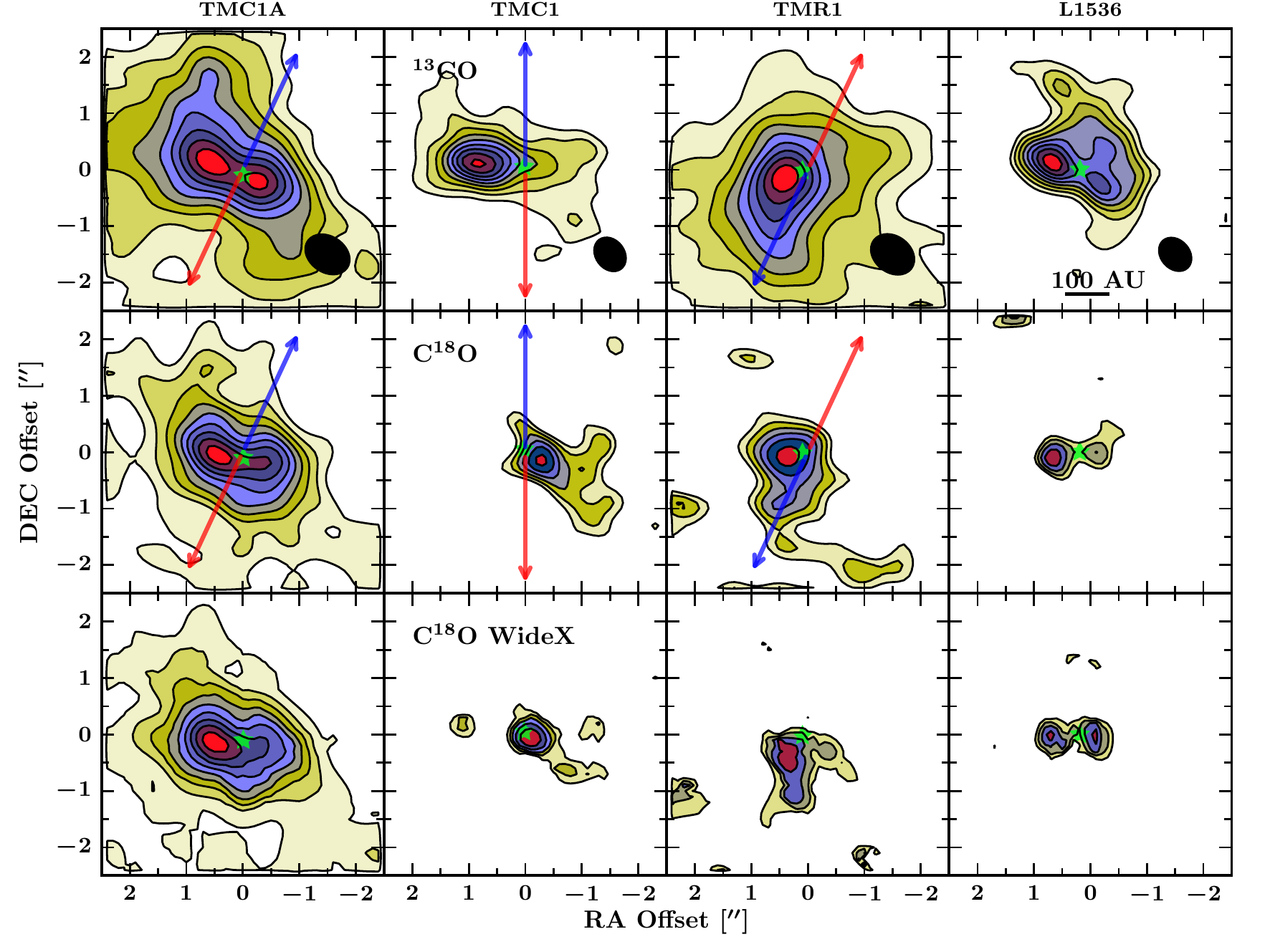}
 \caption{ \tco\ (top) and \ceo\ (bottom) moment zero maps.  The contours are 
drawn at intervals of 10\% of the peak starting from 5$\sigma$.  The 
continuum position is indicated by the green stars and the arrows show the direction 
of the red and blue-shifted outflow components taken from the literature 
\citep[TMC1: 0$^{\circ}$; TMR1: 165$^{\circ}$; TMC1A: 155$^{\circ}$; L1536: 
None][]{ohashi97a, hogerheijde98, brown99}.  The bottom row show \ceo\ 
integrated flux maps from the velocity unresolved spectra (typically 1--2 
channels) taken within the WideX with contours of 10\% of the peak starting from 
3$\sigma$ ($\sim$30 mJy beam$^{-1}$ km s$^{-1}$). The zero moment maps are 
obtained by integrating over the velocity ranges [$\upsilon_{\rm min}, 
\upsilon_{\rm max}$] = [-3, 11], [-1, 11], [-2, 12], [-2, 12] for TMC1A, TMC1, 
TMR1, and L1536, respectively. }
\label{fig:mom0maps}
\end{figure*}

\begin{figure*}[!tbh]
 \centering
  \includegraphics{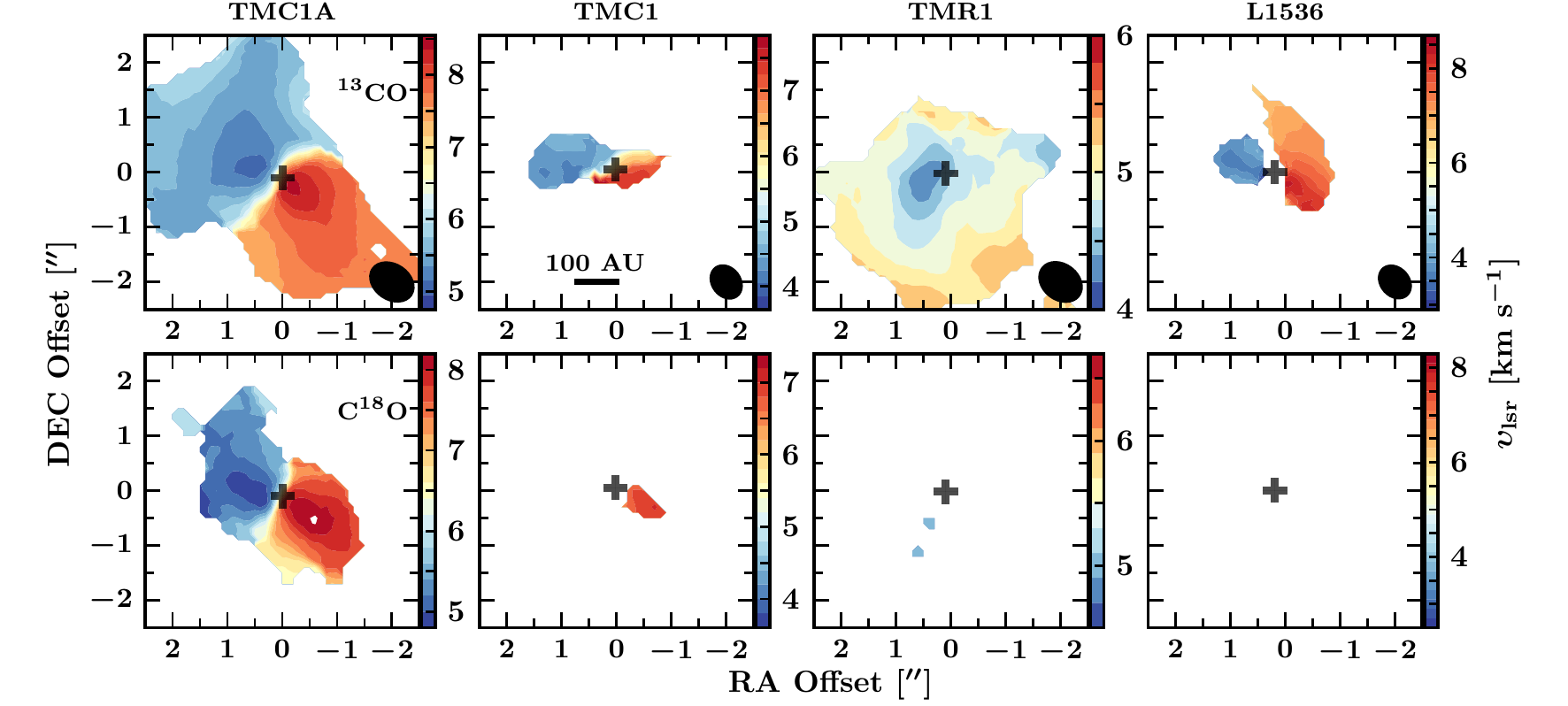}
\caption{\tco\ and \ceo\ moment 1 maps within the inner 5$\arcsec$ (700 AU).  
The solid line indicate the 100 AU scale and the ellipses show the synthesized 
beams.  The gray crosses show the continuum positions. }
\label{fig:mom1maps}
\end{figure*}


The \tco\ and \ceo\ 2--1 lines are detected toward all sources.  
Figure~\ref{fig:lines} presents the spectra integrated within a 1$\arcsec$ 
box around the continuum position.  The integrated line flux densities within 
a 5$\arcsec$ box are tabulated in Table~\ref{tbl:flux}.  The molecular line 
emission is strongest toward TMC1A and weakest toward L1536, while most of the 
emission around the systemic velocity is resolved out.  This can be seen in the 
\tco\ channel map toward TMC1A (Figs.~\ref{fig:lines} and \ref{fig:onemap}, 
see Appendix~\ref{app:data} for channel maps toward other sources) where most of 
the emission is absent near $\upsilon_{\rm lsr} = 6.6$ \kms.  In contrast to \tco, the 
\ceo\ line is detected around the systemic velocities ($\pm 2$ \kms) as shown 
in Fig.~\ref{fig:lines} except toward TMC1A, which shows emission within the 
velocity range comparable to that of \tco.

Single-dish observations of \tco\ $J=$3--2 (except for L1536, which is the 
$J=$2--1 transition) and \ceo\ $J=$2--1 are also shown in Fig.~\ref{fig:lines} 
(right).  The comparison indicates that a large part of the emission at the 
systemic velocity originating from the large-scale envelope is filtered out in 
our PdBI observations. The PdBI spectra within a $1\arcsec$ box around each 
source are also broader than the single-dish spectra.  Thus, our PdBI 
observations are probing the kinematics within the inner few hundreds of AU.

The zeroth moment (i.e., velocity-integrated flux density) maps, are shown in 
Fig.~\ref{fig:mom0maps} toward all four sources.  The morphologies of the \tco\ 
and \ceo\ maps are different toward 3 out of the 4 sources due to \ceo\ emission 
being weaker.  Only toward TMC1A, which has the strongest \ceo\ emission, do the 
\ceo\ and \tco\ show a similar morphologies.  In general, the zeroth moment maps 
indicate the presence of flattened gas structures that are perpendicular to the 
outflow direction except for TMR1.  In most cases, the \ceo\ emission is weak 
and compact toward the continuum position as revealed by our higher S/N 
spectrally unresolved spectra taken with the WideX correlator (bottom panels of 
Fig.~\ref{fig:mom0maps}).

Moment one (i.e., intensity-weighted average velocity) maps are presented in 
Figure~\ref{fig:mom1maps} and show the presence of velocity gradients in the 
inner few hundreds of AU.  These velocity gradients are perpendicular to the 
large-scale outflow.  Large-scale rotation around TMC1A was reported 
previously by \citet{ohashi97a} extending up to 2000 AU.  In addition, 
\citet{hogerheijde98} also detected rotation around TMC1.  The high sensitivity 
and spatial resolution of our observations allow us to map the kinematics on 
much smaller scales and therefore unravel the nature of these gradients.


\section{Analysis} \label{sec:analysis}

\subsection{Continuum: Dust disk versus envelope} \label{sec:diskdust}


\begin{figure*}[!tbh]
 \centering 
\includegraphics{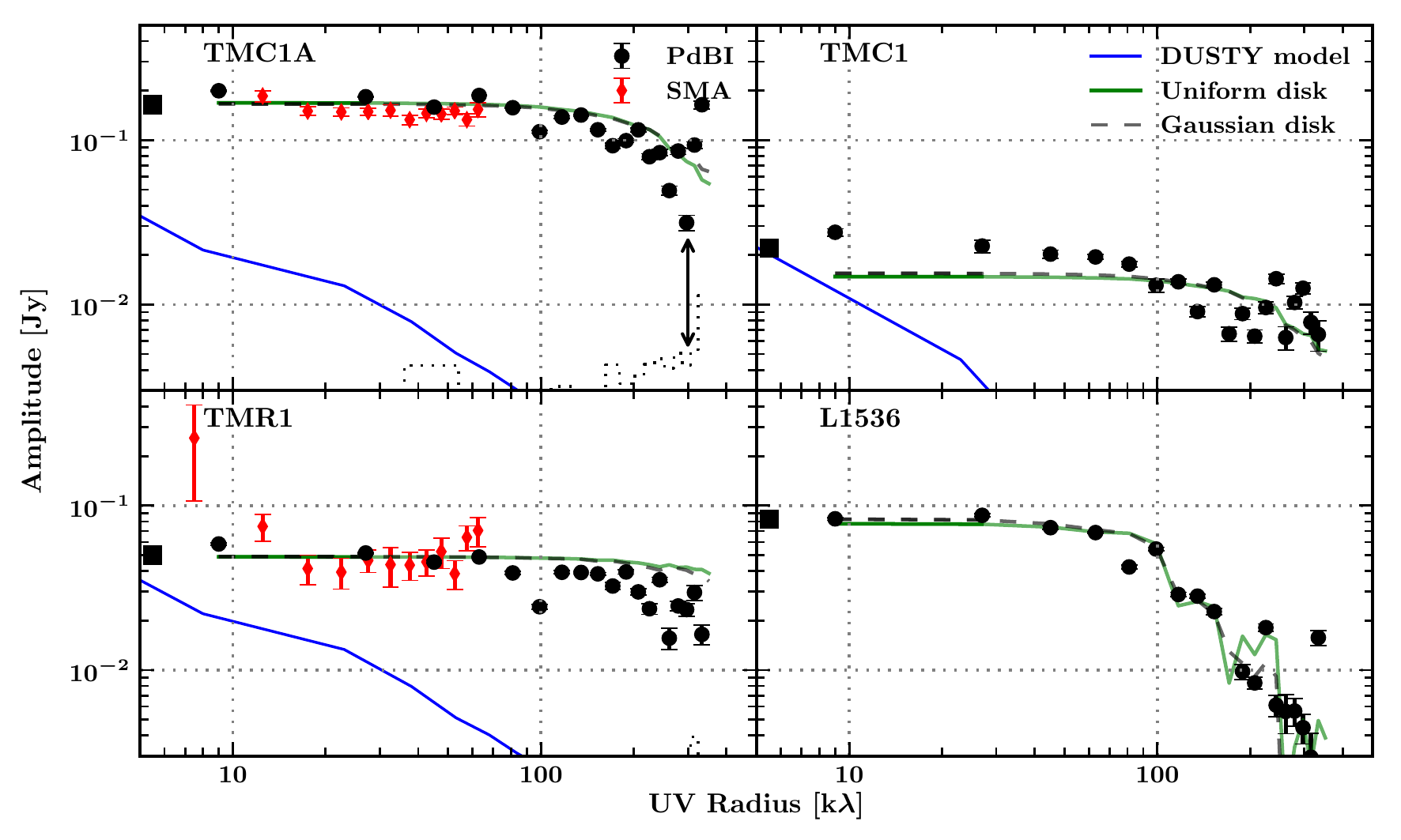}
\caption{Circularly averaged 1.36 mm visibility amplitudes as function of 
projected baselines in k$\lambda$.  The black circles are from this work while 
the red circles are from previous SMA observations at 1.1 mm \citep{prosac09} 
scaled to 1.36 mm with $\alpha = 2.5$.  The error bars indicate the statistical 
errors of each bin.  The blue line is the simulated continuum emission 
of the spherical envelope model from \citet{kristensen12} sampled using the 
observed $uv$ points.  An arrow in the first panel shows an example of the 
continuum excess toward TMC1A.  Solid squares are the 1.36 mm single-dish flux.  
The zero-expectation level given in dotted line is the expected amplitude due to 
noise in the absence of source emission. }
\label{fig:uvamp}
\end{figure*}


The advantage of an interferometric observation is that one can disentangle the 
compact disk-like emission from the large-scale envelope.  Thus, an estimate of 
the disk mass can be obtained using both the single-dish and the spatially 
resolved continuum observations (visibility amplitudes).  Figure~\ref{fig:uvamp} 
shows the circularly averaged visibility amplitudes as a function of the projected 
baselines in k$\lambda$ (assuming spherically symmetric emission).  All sources 
show strong emission on long baselines that is near constant with increasing 
baseline.  The continuum visibilities of TMR1 and TMC1A are consistent with the 
scaled ($F_{\nu} \propto \nu^{\alpha}$) 1.1 mm SMA observations (red circles, 
\citealt{prosac09}) with $\alpha = 2.5$.  An excess of continuum emission at the 
longest baselines compared with the expected noise is seen toward all sources, 
which corresponds to unresolved continuum structure at scales smaller than 320 
k$\lambda$ (50 AU radius).  Moreover, rapidly increasing amplitudes at short 
baselines are typically associated with the presence of a large-scale envelope.  
This combination of features in the visibilities is a typical signature of 
embedded disks \citep[e.g.,][]{looney03, jorgensen05b, enoch09}.

The first important step is quantifying the large-scale envelope contribution to 
the $uv$ amplitudes.  Using radiative transfer models, \citet{prosac09} 
estimated that a spherical envelope contributes at most 4\% at 50 k$\lambda$ at 
1.1 mm, which translates to 2\% at 1.36 mm.  By using the single-dish 850 
$\mu$m ($S_{\rm 15 \arcsec}$) and the 50 k$\lambda$ fluxes ($S_{\rm 50 
k\lambda}$), the envelope, $S_{\rm env}$, and disk, $S_{\rm disk}$, flux 
densities can be estimated by solving the following system of equations (Eqs. 2 
and 3 in \citealt{prosac09}):
\begin{eqnarray}
 S_{\rm 50 k \lambda} & =  & S_{\rm disk} + c \times S_{\rm env} \\
 S_{\rm 15 \arcsec} &  = & S_{\rm disk} (1.36/0.85)^{\alpha} + S_{\rm env}.
\end{eqnarray}
We have adopted $\alpha \simeq 2.5$ to translate the 850 $\mu$m disk flux 
to 1.36 mm \citep{prosac07} and $c = 0.02$ as the fractional envelope 
contribution.  The derived disk masses within 100 AU diameter are tabulated in 
Table~\ref{tbl:sources} using the OH5 \citep{OH94} opacities at 1.36 mm and a 
dust temperature of 30 K.  Our values for TMR1 and TMC1A are consistent with 
\citet{prosac09}.  The envelope masses listed in Table~\ref{tbl:sources} were  
estimated using Equation (1) of \citet{prosac09}.  Three of the sources (TMC1, 
TMR1, TMC1A) have similar envelope masses of $\sim 0.1 \ M_{\odot}$, which are 
higher than the disk masses.  In contrast, at most a tenuous envelope seems to 
be present around L1536.  Using the disk to envelope mass ratio as an 
evolutionary tracer, L1536 is the most evolved source in our sample, which is 
consistent with its high bolometric temperature.  Thus, these sources are indeed 
near the end of the main accretion phase where a significant compact component 
is present in comparison to their large-scale envelope.

In interferometric observations toward YSOs, the lack of short baselines 
coverage may result in underestimation of the large-scale emission and, 
consequently, overestimating the compact structure.  \citet{kristensen12} 
calculated 1D spherically symmetric models derived using the 1D continuum 
radiative transfer code DUSTY \citep{dusty} for a number of sources, 
including TMC1A, TMR1 and TMC1. The model is described by a power law density 
structure with an index $p$ ($n = n_0 (r /r_0 )^{-p}$) where $n$ is the number 
density and with an inner boundary $r_{0}$ at $T_{\rm dust} = 250$ K.  The 
parameters are determined through fitting the observed spectral energy 
distribution (SED) and the 450 and 850 $\mu$m dust emission profiles.  These 
parameters are tabulated in Table~\ref{tbl:1dmodels}.  The models in 
\citet{kristensen12} do not take into account the possible contribution by the 
disk to the submillimeter fluxes and the envelope masses are consequently higher 
than ours by up to a factor of 2.  Using these models, the missing flux can be 
estimated by modelling the large-scale continuum emission as it would be 
observed by the interferometer (blue line in Fig.~\ref{fig:uvamp}).  We can 
therefore place better constraints on the compact flattened structure.


\subsection{Constraining the dust disk parameters} \label{sec:diskparams}

\begin{table}[htb!]
 \centering
 \caption{Disk sizes and masses derived from continuum visibilities modelling 
using uniform, Gaussian and power-law disk models.}
 \label{tbl:udgdtbl}
\begin{tabular}{ c c c c c}\toprule \hline
    Object & \multicolumn{4}{c}{Parameters} \\
    & \multicolumn{1}{c}{$R_{\rm disk}$} & $M_{\rm disk}$ & $i$ & PA\\
    &  [AU] & [$10^{-3}$ M$_{\odot}$] & [$^{\circ}$] & [$^{\circ}$] \\
    \hline
    & \multicolumn{4}{c}{Uniform disk} \\
    TMC1A & 30 -- 50 & 48$\pm 6$ & 20--46 & 60--99 \\
    TMC1 &  7 -- 124 & 4.7$^{+4}_{-2}$ & 30--70 & 65--105 \\
    TMR1 &  7 -- 40 & 15$\pm 2$ & 5--30 &  -15--15 \\
    L1536 & 100 -- 171 & 22$\pm 5$ & 30--73 & 60--100\\
    \hline
    & \multicolumn{4}{c}{Gaussian disk} \\
    TMC1A & 17--32 & 48$\pm 6$ & 24--56 & 62--96 \\
    TMC1 & 7--80 & 4.5$^{+4}_{-2}$ & 30--70 & 65--105 \\
    TMR1 & 7--20 & 15$\pm 2$ & 5--30  & -15--15 \\
    L1536 & 53--112 & 24$\pm 4$ & 33--75 & 60--100 \\
    \hline
    & \multicolumn{4}{c}{Power-law disk} \\
    TMC1A & 80--220 & 41$\pm 8$ & 24--76 & 60--99 \\
    TMC1 & 41--300 & 5.4$^{+20}_{-0}$ & 20--80 & 55--115 \\
    TMR1 & 7--55 & 10$^{+100}_{-9} $ & 2--45  & -25--40 \\
    L1536 & 135--300 & 19$^{+18}_{-4}$ & 20--56 & 50--68 \\    
   \bottomrule
   \end{tabular}
\end{table}

\begin{figure}[!tbh]
 \centering
 \includegraphics{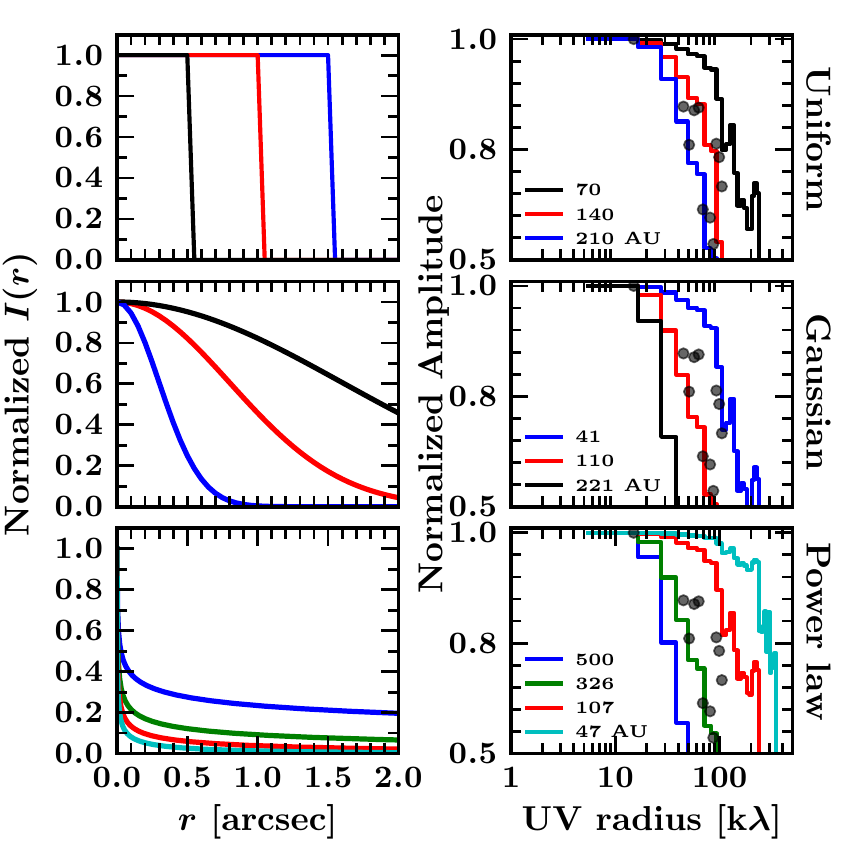}
 \caption{Left: Normalized intensity profiles for the three different continuum 
models: Uniform disk (top), Gaussian Disk (middle), and power-law disk 
(bottom). Right: Normalized visibilities as function of $uv$ radius.  For 
comparison, the L1536 binned data are shown in gray circles.}
\label{fig:modelvis}
\end{figure}


The goal of this section is to estimate the size and mass of the compact 
flattened dust structure or the disk using a number of more sophisticated 
methods than Eq.~(1) and (2) to compare with the estimates from the gas in the 
next section.  We define the disk component to be any compact flattened 
structure that deviates from the expected continuum emission due to the 
spherical envelope model.  In the following sections, the disk sizes will be 
estimated by fitting the continuum visibilities with the inclusion of emission 
from the large-scale structure described above.  As the interferometric 
observation recovers a large fraction of the single-dish flux toward L1536, its 
large-scale envelope contribution is assumed to be insignificant, and therefore 
not included.  The best-fit parameters are found through $\chi^2$ minimization 
on the binned amplitudes.  The range of values of $R_{\rm disk}$, $M_{\rm 
disk}$, $i$ and $PA$ shown in Table~\ref{tbl:udgdtbl} correspond to models with 
$\chi^2$ between $\chi_{\rm min}^{2}$ and $\chi_{\rm min}^2 + 15$.


\subsubsection{Uniform disk}\label{sec:uniformdisks}

The first estimate on the disk size and mass is obtained by fitting a uniform 
disk model assuming optically thin dust emission \citep[see 
Fig.~\ref{fig:modelvis},][]{eisner12}.  Uniform disks are described by a 
constant intensity $I$ within a diameter $\theta_{\rm ud}$ whose flux 
is given by $F = \int I \cos \theta \ d\Omega = I \pi \left ( \frac{\theta_{\rm 
ud}}{2} \right )^2$.  The visibility amplitudes are given by: 
\begin{equation}
 V(u,v) = F \times 2 \frac{J_1 \left ( \pi \theta r_{uv} \right )}{\pi \theta 
r_{\rm uv}},
\end{equation}
where $J_1$ is the Bessel function of order 1 and $r_{\rm uv}$ is the 
projected baseline in terms of $\lambda$.  Deprojection of the 
baselines follow the formula given in \citet{berger07}:
\begin{eqnarray}
 u_{\rm PA} & = & u \ \cos PA + v \ \sin PA, \\
 v_{\rm PA} & = & -u \ \sin PA + v \ \cos PA ,\\
 r_{uv, i} & = & \sqrt{(u_{\rm PA}^2 + v_{\rm PA}^2 \ \cos(i)^2)},
\end{eqnarray}
where $PA$ is the position angle (East of North) and $i$ is the inclination.

The best fit parameters are tabulated in Table~\ref{tbl:udgdtbl} with disk 
masses estimated using a dust temperature of 30 K and OH5 opacities  
\citep[Table 5 of][]{OH94} within $\frac{\theta_{\rm ud}}{2}$ radius.  The disk 
radii vary between 7 to 171 AU.  The smallest disk size is found toward TMR1, 
which suggests that most of its continuum emission is due to the large-scale 
envelope.  Moreover, disk sizes around TMR1 and TMC1A are 65\%--90\% lower 
than those reported by \citet[][Table 3]{eisner12} since we have included the 
spherical envelope contribution.  This illustrates the importance of including 
the large-scale contribution in analyzing the compact structure.


\subsubsection{Gaussian disk}\label{sec:gaussdisks}

The next step is to use a Gaussian intensity distribution, which is a 
slightly more realistic intensity model that represents the mm emission 
due to an embedded disk.  The visibility amplitudes are described by 
the following equation: 
\begin{equation}
 V(u,v) = F \times \exp \left ( - \frac{ \pi \theta_{\rm G} r_{uv} }{4 \log 2} 
\right ),
\end{equation}
where  $\theta_{\rm G}$ is the $FWHM$ of the Gaussian distribution and 
$F$ is the total flux.

The difference between the Gaussian fit presented in Table~\ref{tbl:sources} 
and this section is the inclusion of the large-scale emission.  In 
Table~\ref{tbl:sources}, the Gaussian fit gives an estimate of the size of 
emission in the observed total image, while this section accounts for the 
simulated large-scale envelope emission in order to constrain the compact 
flattened structure.  The free-parameters are similar to the uniform disk except 
that the size of the emitting region is defined by full-width half maximum 
($FWHM$).  For a Gaussian distribution, most of the radiation (95\%) is emitted 
from within $2\sigma_G \sim 0.85 \times FWHM$, thus we define the disk radius to 
be $0.42 \times FWHM$.  The best-fit parameters are tabulated in 
Table~\ref{tbl:udgdtbl} and the corresponding emission is included in 
Fig.~\ref{fig:uvamp}.  The resulting disk sizes are slightly lower than those 
derived using a uniform disk.

The visibility amplitudes in Fig.~\ref{fig:uvamp} show the presence of an 
unresolved point source component at long baselines indicated by the arrow in 
the first panel.  \citet{jorgensen05b} noted that a three components model 
(large-scale envelope + Gaussian disk + point-source) fits the continuum 
visibilities.  By using such models, the disk sizes generally increase by 20--30 
AU with the addition of an unresolved point flux, which is comparable to the 
uniform disk models.  Such models were first introduced by \citet{mundy96} 
since the mm emission seems to be more centrally peaked than a single Gaussian. 
  The unresolved point flux is more likely due to the unresolved disk structure 
since the free-free emission contribution is expected to be low toward Class I 
sources \citep{hogerheijde98}.


\subsubsection{Power-law disk}\label{sec:powlawdisks}

The next step in sophistication is to fit the continuum visibilities with a 
power-law disk structure.  The difference is that the first two disk models are 
based on the expected intensity profile of the disk, while the power-law disk 
models use a more realistic disk structure.  Similar structures were previously 
used by \citet[][see also \citealt{lay94, mundy96, dutrey07, malbet05}]{lay97}. 
 We adopt a disk density structure described by $\Sigma_{\rm disk} = \Sigma_{\rm 
50 AU} \left ( R / 50 {\rm AU }\right)^{-p}$ where $\Sigma_{\rm 50\ AU}$ is the 
surface density at 50 AU with a temperature structure given as $T_{\rm disk} = 
1500 \left (R/0.1 {\rm AU} \right )^{-q}$.  The inner radius is fixed at 0.1 AU 
with a dust sublimation temperature of 1500 K, while the disk outer radius is 
defined at 15 K.  The free parameters are $p$, $q$, $\Sigma_{\rm 50 AU}$, $i$ 
and $PA$ where $i$ is the inclination with respect of the plane of the sky in 
degrees (0$^{\circ}$ is face-on) and $PA$ is the position angle (East of North). 
The visibilities are constructed by considering the flux from a thin ring given 
by:
\begin{equation}
dS = \frac{2 \pi \ \cos \ i }{d^2} B_{\nu}(T) \left (1 - \exp^{-\tau} \right ) 
 R \ dR,
\end{equation}
where $d$ is the distance, $i$ is the inclination ($0^{\circ}$ is face-on), 
$\tau = \frac{\Sigma \left ( R \right ) \kappa}{\cos \ i}$ is the optical depth 
and $B_{\nu} (T)$ is the Planck function.  For the dust emissivity ($\kappa$), 
we adopt the OH5 values.  Since the model is axisymmetric, the visibilities 
are given by a one-dimensional Hankel transform which is given by the analytical 
expression:
\begin{equation}
V \left ( u, v \right) = dS \times J_0\left ( 2 \pi R r_{uv} \right ),
\end{equation}
where $J_0$ is the Bessel function of order 0.  The total amplitude at a given 
$r_{\rm uv}$ is integrated over the whole disk.

The first advantage of such modelling is the treatment of the optical depth 
which gives better mass approximations.  Secondly, the unresolved 
point-flux is already included in the model by simulating the emission from 
the unresolved disk component.  Figure~\ref{fig:modelvis} shows the 
normalized visibilities for a fixed $p=0.9$ and a number of $R_{\rm disk}$.  
The best-fit parameters are tabulated in Table~\ref{tbl:udgdtbl}.

The derived disk masses obtained by integrating out to $\sim 50$ AU radius are 
similar to those from the 50 k$\lambda$ point (Table~\ref{tbl:sources}).  This 
is expected as the 1.3 mm continuum emission ($\kappa_{\rm OH5} = 0.83$ 
cm$^{2}$g$^{-1}$) is on average nearly optically thin \citep{prosac07}.  The 
optical depth can be up to $0.4$ at radii $< 10$ AU for the best-fit power-law 
disk models with $i > 70^{\circ}$.  A wide range of `best-fit' masses is found 
toward TMR1, which is mainly due to the uncertainties in the inclination.

More recently, \citet{eisner12} presented embedded disk models derived from 
continuum data by fitting the $I$ band image, SED and 1.3 mm visibilities.  Our 
continuum disk radii toward TMC1A and TMR1 are consistent with their results. 
However, the high quality of our data allow us to rule out disks with $R_{\rm 
out}>100$ AU toward TMR1.  On the other hand, our disk radius toward L1536 is a 
factor of 3 higher than that reported in \citet{eisner12} due to the difference 
in the treatment of the large-scale envelope.

In summary, different methods have been used to constrain the disk parameters 
from thermal dust emission.  The first two models, Gaussian and uniform disk 
models, are based on the intensity profile, while the power-law disk models use 
a simplified disk structure with given density and temperature profiles.  The 
intensity based models find the smallest disk sizes, while the power-law disk 
models predict up to a factor of two larger disk sizes.  On the other hand, the 
disk masses from these fitting methods are similar.  The advantage of the 
power-law disk modeling is the determination of the density and temperature 
structures including the unresolved inner part of the disk.  The continuum 
images provide no kinematic information, however, so the important question is 
whether these flattened structures are rotationally supported disks and if so, 
whether the continuum sizes agree with those of RSDs.


\subsection{Line analysis: Keplerian or not?}\label{sec:keplerian?}

The nature of the velocity gradient can already be inferred from the moment 
maps.  First, the \tco\ integrated flux maps shown in Fig.~\ref{fig:mom0maps} 
indicate the presence of flattened gas structures perpendicular to the outflow 
direction.  Second, the velocity gradient is also oriented perpendicular to the 
outflow within hundreds of AU, which is a similar size scale to the compact dust 
structure measured in Section~\ref{sec:diskparams}.  On the other hand, the 
velocity gradient does not show a straight transition from the red to the blue 
shifted component as expected from a Keplerian disk.  Taking TMC1A as an 
example, the blue-shifted emission starts at the North side and skews inward 
toward the continuum position.  Such a skewed moment 1 map is expected for a 
Keplerian disk embedded in a rotating infalling envelope \citep{sargent87, 
brinch08}.  Thus, the moment maps indicate the presence of a rotating infalling 
envelope leading up to a rotationally dominated structure.

In image space, position velocity (PV) diagrams are generally used to determine 
the nature of the velocity gradient.  Analysis of such diagrams starts by 
dividing it into four quadrants around the $\upsilon_{\rm lsr}$ and source 
position.  A rotating structure occupies two of the four quadrants that are 
symmetric around the center, while an infalling structure will show emission in 
all quadrants \citep[e.g.,][]{ohashi97a, brinch07}.  In addition, one can infer 
the presence of outflow contamination by identifying a velocity gradient in PV 
space along the outflow direction \citep[e.g.,][]{cabrit96}.


\begin{figure*}[!tbh]
 \centering
 \includegraphics{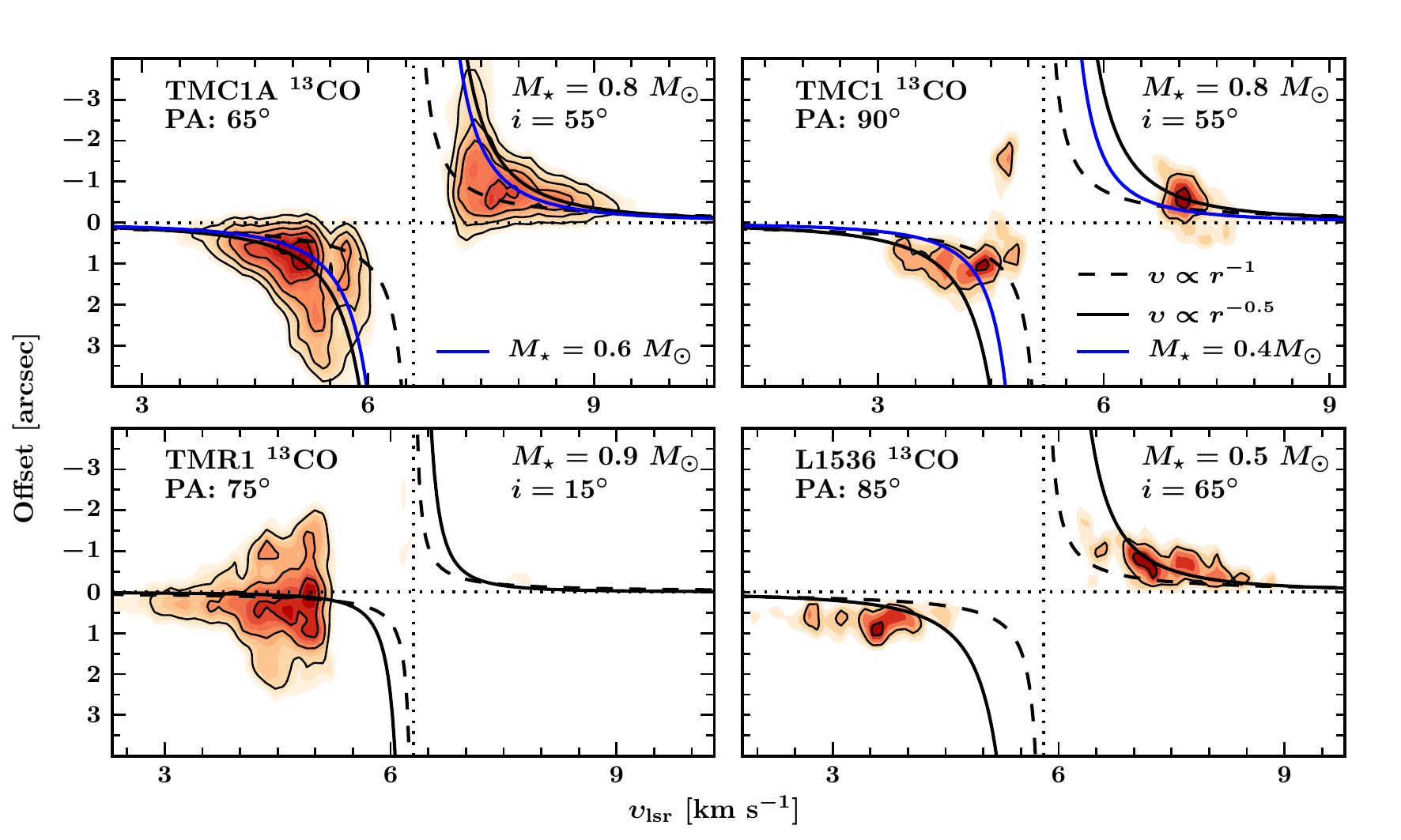}
\caption{\tco\ position velocity diagrams along the direction perpendicular to 
the outflow: TMC1A (top left), TMC1 (top right), TMR1 (bottom left), and L1536 
(bottom right).  The color contours start from 3$\sigma$ and increase by 
1$\sigma$ while the black line contours are drawn at 5$\sigma$, 8$\sigma$, 
11$\sigma$, and so on up to the peak intensity.  The black solid curves show the 
best-fit Keplerian curves, while the dashed lines indicate the $r^{-1}$ curves.  
For TMC1A and TMC1 (top row), the blue solid line presents a $0.4 \ M_{\odot}$ 
Keplerian curve. }
\label{fig:pvdiags}
\end{figure*}


Figure~\ref{fig:pvdiags} presents PV diagrams at a direction perpendicular to 
the outflow direction; for L1536, the direction of elongation of the continuum 
was taken.  In general, the PV diagrams are consistent with Keplerian profiles, 
except for TMR1.  The PV diagram toward TMR1 seems to indicate that the \tco\ 
line is either dominated by infalling material or the source is oriented toward 
us, which is difficult to disentangle.  However, recent \mco\ 3--2 and 6--5 
maps presented in {Y{\i}ld{\i}z et al}, (in prep.) suggest that the disk is more 
likely oriented face-on ($i < 15^{\circ}$).  Furthermore, a velocity gradient is 
present in the \tco\ PV analysis both along and perpendicular to the outflow 
direction indicating minor outflow contamination for this source.  

The focus of this paper is to differentiate between the infalling gas and the 
rotationally supported disk.  Such analysis in image space is not sensitive to 
the point where the infalling rotating material ($\upsilon \propto r^{-1}$) 
enters the disk and becomes Keplerian ($\upsilon \propto r^{-0.5}$).  Thus, 
additional analysis is needed to differentiate between the two cases.

As pointed out by \citet{lin94}, infalling gas that conserves its angular 
momentum exhibits a steeper velocity profile ($\upsilon \propto r^{-1}$) than 
free-falling gas ($\upsilon \propto r^{-0.5}$).  In the case of spectrally 
resolved optically thin lines, the peak position of the emission corresponds to 
the maximum possible position of the emitting gas \citep{sargent87, harsono13a}. 
 On the other hand, molecular emission closer to the systemic velocity is 
optically thick and, consequently, the inferred positions are only lower 
limits.  Thus, with such a method, one can differentiate between the Keplerian  
disk and the infalling rotating gas.  Moreover, \citet{harsono13a} suggested 
that the point where the two velocity profiles meet corresponds to the size of 
the Keplerian disk, $R_{\rm k}$.  In the following sections, we will attempt to 
constrain the size of the Keplerian disk using this method, which we will term 
the $uv$-space PV diagram.

The peak positions are determined by fitting the velocity resolved visibilities 
with Gaussian functions \citep{lommen08, prosac09}.  It can be seen from the 
channel maps (Fig.~\ref{fig:onemap} and Appendix A) that a Gaussian brightness 
distribution is a good approximation in determining the peak positions of the 
high velocity channels.

To characterize the profile of the velocity gradient, the peak positions are 
projected along the velocity gradient.  This is done by using the following 
transformation 
\begin{eqnarray}
 x_{PA} & = & x \cos(PA) + y \sin(PA), \\
 y_{PA} & = & -x \sin(PA) + y \cos(PA), 
\end{eqnarray}
where $x,y$ are the peak positions and $x_{\theta}, y_{\theta}$ are the 
rotated positions.  A velocity profile ($\upsilon \propto r^{-\eta}$) is then 
fitted to a subset of the red- and blue-shifted peaks to determine the best 
velocity profile.

\subsubsection{TMC1A}

\begin{figure}[bthp]
 \centering
 \includegraphics{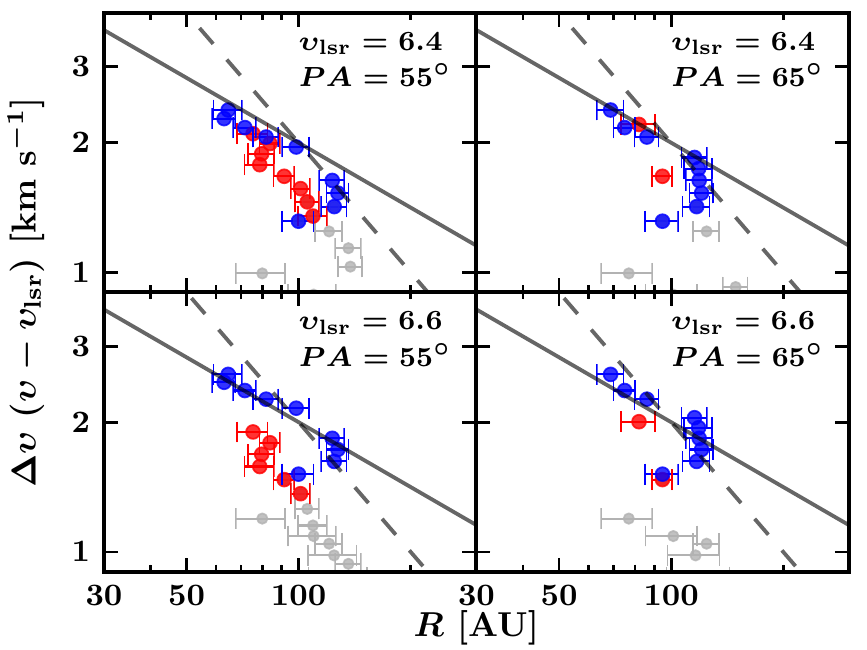}
 \caption{\ceo\ $uv$-space PV diagrams toward TMC1A with $PA = 55^{\circ}, 
65^{\circ}$ and $\upsilon_{\rm lsr} = 6.6, 6.4$ \kms.  The solid lines are the 
$r^{-0.5}$ curves and dashed lines are the $r^{-1}$ curves.  The red and blue 
colors indicate the red- and blue-shifted components, respectively.  Gray 
points show the rest of the peaks that were not included in the fitting.}
\label{fig:tmc1avelprof1}
\end{figure}

\begin{figure*}[htbp]
 \centering
 \includegraphics{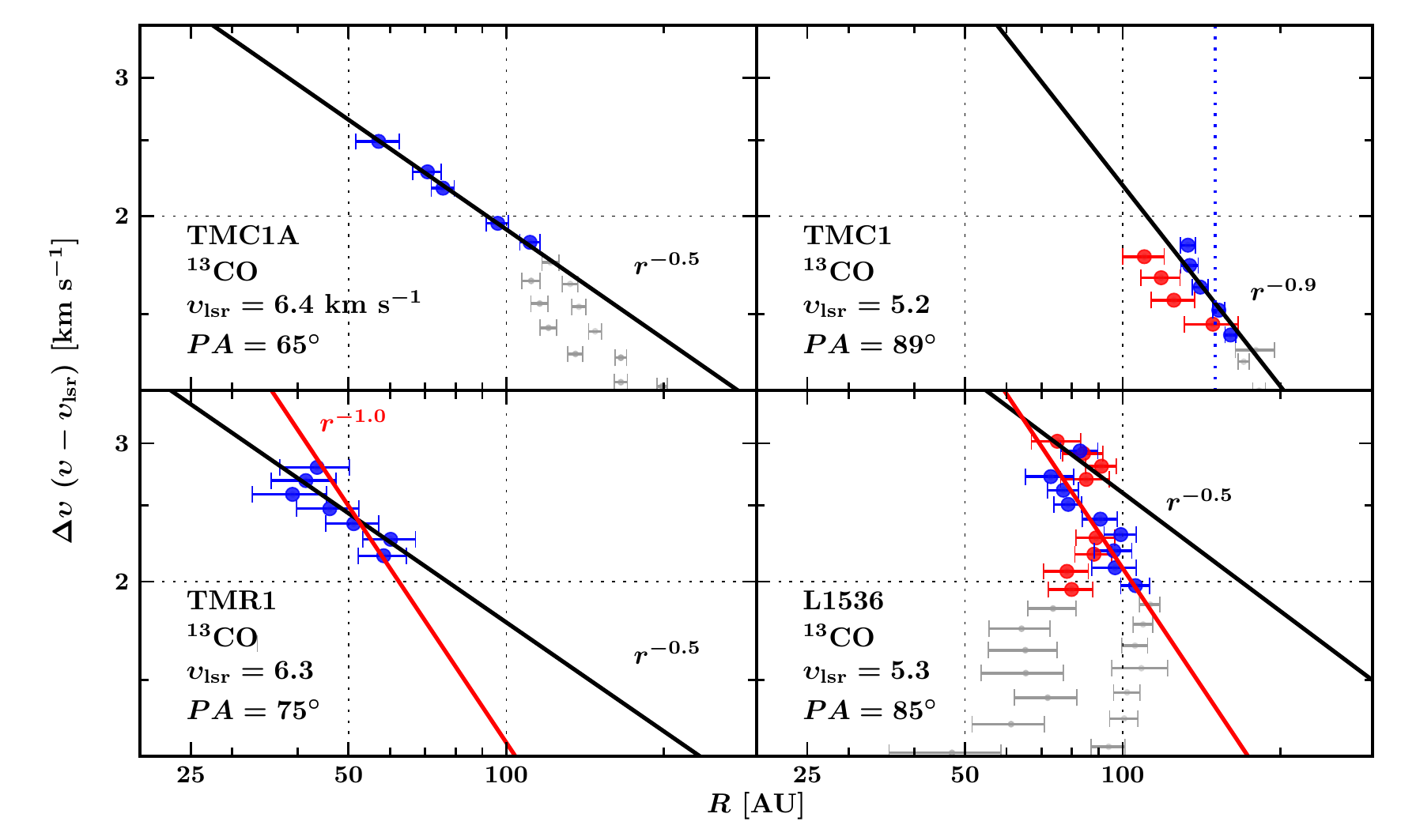}
 \caption{\tco\ $uv$-space PV diagrams toward all sources with the best fit 
profiles superposed in black lines (except for L1536).  The red lines indicate a 
$\upsilon \propto r^{-1}$ as comparison.  Gray points show the peaks that were 
not included in the fitting. }
 \label{fig:all13coprofs}
\end{figure*}

TMC1A shows the greatest promise so far for an embedded Keplerian disk around a
Class I protostar .  The moment 1 maps (Fig.~\ref{fig:mom1maps}) indicate a 
gradual change from blue- to red-shifted gas in the inner few hundreds of AU, 
which is consistent with a Keplerian disk.  Figure~\ref{fig:tmc1avelprof1} 
presents the \ceo\ velocity profiles using the nominal $\upsilon_{\rm lsr}$ and 
$PA$ perpendicular to the outflow direction along with small deviations from 
those values, which correspond to the uncertainty in outflow direction.  In the 
absence of foreground clouds, one expects that the blue and red-shifted peaks 
overlap for a rotating system, which is the case if one adopts $\upsilon_{\rm 
lsr} = +6.4$ \kms.  The \ceo\ and C$^{17}$O observations presented in 
\citet{jorgensen02} indicate a $\upsilon_{\rm lsr} = +6.6$ \kms, however 
$\upsilon_{\rm lsr} = +6.4$ \kms\ is consistent with the N$_2$H$^{+}$ 
observations toward the nearby L1534 core \citep{caselli02}.  For comparison, 
the \tco\ velocity profile with $\upsilon_{\rm lsr} = +6.4$ \kms\ and $PA$ 
perpendicular to the outflow direction is shown in Fig~\ref{fig:all13coprofs}.  
Both of the \tco\ and \ceo\ lines show a velocity profile close to $\upsilon 
\propto r^{-0.5}$, which indicates that the gas lines trace similar 
rotationally supported structures.  The typical uncertainty in the power-law 
slope is $\pm 0.2$.  In the region where the red- and blue-shifted peaks 
overlap, a clear break from $r^{\sim -0.5}$ to a steeper slope of $r^{-1}$ can 
be seen at $\sim$100 AU.  This break may shift inward at other $PA$s and 
$\upsilon_{\rm lsr}$ to 80 AU.  However, it is clear that the flattened 
structure outside of 100 AU flattened structure is dominated by infalling 
rotating motions.  Thus, the embedded RSD toward TMC1A has a radius between 
80--100 AU.

To derive the stellar mass associated with the rotationally dominated region, 
simultaneous fitting to the PV diagrams in both image and  $uv$ space has 
been performed.  The best-fit stellar mass associated with the observed 
rotation toward TMC1A is 0.53$^{+0.2}_{-0.1}$ $M_{\odot}$ at $i = 55^{\circ} 
\pm 10^{\circ}$.

Additional analysis is performed for TMC1A with a simple radiative transfer 
model to confirm the extent of the Keplerian disk following a method similar to 
that used by \citet{murillo13b} for the VLA 1623 Class 0 source.  Such 
fitting is only performed for this source to investigate whether they give 
similar results.  A flat disk model given by:
\begin{eqnarray}
T & = & T_0 \left ( r / r_0 \right )^{-q}, \\
N & = & N_0 \left ( r / r_0 \right )^{-p},
\end{eqnarray}
where $N$ is the gas column density with $T_0$ and $N_0$ as the reference gas 
temperature and column densities, respectively, at the reference radius, $r_0$. 
 Similar to the power-law disk models in Section~\ref{sec:powlawdisks}, we have 
used $T_0 = 1500$ K at $r_0$ of 0.1 AU.  The images are then convolved with the 
clean beams in Table~\ref{tbl:weights}.  We have assumed $i = 55^{\circ}$ 
and $PA = 65^{\circ}$.  For such orientation, we find that a Keplerian 
structure within the inner 100 AU is needed, surrounded by a rotating infalling 
envelope.  The radius of the Keplerian structure is similar to the radius 
derived from previous methods ($\sim 100$ AU), however, a stellar mass of $0.8 
\pm 0.3 \ M_{\odot}$ is needed to reproduce the observed molecular lines.  This 
is consistent with \citet{yen13} calculated at $i = 30^{\circ}$, but a factor 
of 1.5 higher than the value obtained earlier (although the two values agree 
within their respective uncertainties).

\subsubsection{TMC1}
\begin{figure}
\centering
\includegraphics{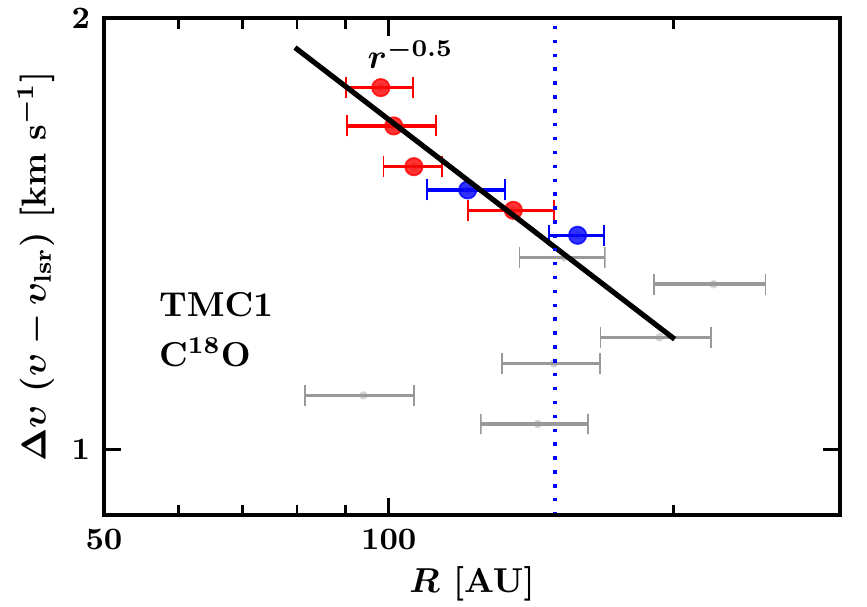}
\caption{\ceo\ $uv$-space PV diagrams toward TMC1 derived from the emission 
peaks.  The solid line indicates the best-fit power-law slope to the peaks 
between 1.3--1.9 \kms.  The vertical blue dotted lines indicate the 150 AU 
radius.  The red and blue colors indicate the red- and blue-shifted components, 
respectively.}
\label{fig:tmc1velprof}
\end{figure}

The velocity profile for TMC1 projected along $PA = 90^{\circ}$ is shown in 
Figure~\ref{fig:all13coprofs} for $^{13}$CO.  The moment maps toward this source 
indicate the presence of a compact flattened structure around the continuum 
position.  Similar to TMC1A, the PV diagram also indicates a rotationally 
dominated structure.  However, in contrast to TMC1A, the \tco\ line is best 
described by an $r^{-1}$ profile indicating that it traces infalling rotating 
gas, while the \ceo\ line is fitted well with a $r^{-0.5}$ slope as shown in 
Fig.~\ref{fig:tmc1velprof} indicative of an RSD.  Such discrepancy can occur if 
the two lines are tracing different regions.  The moment zero maps in 
Fig.~\ref{fig:mom0maps} already gave some indication that the \tco\ and \ceo\ 
emissions do not arise from similar structure.  Furthermore, the disk mass 
toward TMC1 as inferred through the continuum is the lowest in our sample, 
hence it is also possible that the embedded RSD is not massive enough to 
significantly contribute to both the \tco\ and \ceo\ emission.  The relatively 
more optically thick \tco\ traces the inner envelope and thus is dominated by 
its kinematics, while the rotationally supported structure significantly 
contributes to the \ceo\ line.  A smooth extrapolation of the envelope 
down to 100--150 AU scales predicts \ceo\ emission $< 18\%$ of the observe 
emission at $> 1.5$ \kms, suggesting that the majority of the emission at these 
velocities originates in a disk (see also models by \citealt{harsono13a}, 
figure 8).   Hence, using the \ceo\ line, the fit indicates the presence of 
an RSD extending between 100--150 AU in radius.  The best-fit stellar mass is 
0.54$^{+0.2}_{-0.1}$ $M_{\odot}$ at $i=55^{\circ}$ with typical error of $\pm 
0.2$ \kms\ on the determination of $\upsilon_{\rm lsr}$.   The stellar mass is a 
factor of 1.5 lower than that previously derived by \citet{hogerheijde98} as 
indicated by the black lines in Fig.~\ref{fig:pvdiags}.  However, a stellar mass 
of 0.4 $M_{\odot}$ also reasonably fits the \tco\ PV diagram.  In addition, the 
discrepancy between the \tco\ and \ceo\ velocity profiles may suggest that the 
most stable RSD is less than the \ceo\ break point.  Thus, we will use 100 AU as 
the size of the RSD toward TMC1.

\subsubsection{TMR1}

The determination of the velocity profile toward TMR1 is difficult.  Adopting 
$\upsilon_{\rm lsr} = +6.3$ \kms\ \citep{kristensen12, yildiz13a},  the 
red-shifted peaks do not coincide with the blue-shifted peaks as expected from 
a symmetric rotating infalling structure in the absence of foreground clouds. 
 The $^{13}$CO velocity profile is uncertain with a slope between 0.5 to 1 
(Fig.~\ref{fig:all13coprofs}).  However, a velocity gradient is present in the 
PV analysis along and perpendicular to the outflow direction, which suggest 
outflow contamination toward TMR1.  With the $5\sigma$ threshold used for this 
analysis, only 3 peaks are available in the \ceo\ line, which are located along
the blue outflow lobe as inferred from the \mco\ $J=$3--2 and $J=$6--5 maps in 
Y{\i}ld{\i}z et al, in prep.  The on-source \ceo\ line is detected within the 
WideX correlator, which is double peaked within $100$ AU diameter 
(Fig.~\ref{fig:mom0maps}).   Thus, the non-detection of the turn-over from 
$r^{-1}$ to $r^{-0.5}$ indicates that the disk is much smaller than the 
synthesized beam ($<0.7\arcsec = 98$ AU diameter).  Also, the RSD has to be 
less massive than TMC1 and TMC1A due to the lack of on-source \ceo\ emission 
and rotational motion.  Thus, the compact flattened dust structure at $>50$ 
k$\lambda$ toward this source cannot be fully associated with an RSD.  The 
molecular emission from TMR1 must be mostly due to the infalling rotating 
envelope and the entrained outflow gas similar to the conclusion of 
\citet{prosac09}.

\subsubsection{L1536}

Figure~\ref{fig:all13coprofs} shows the $uv$-space PV diagram with systemic 
velocity of $\upsilon_{\rm lsr} = +5.3$ \kms.  The peak positions were shifted 
to the center of the two peaks seen in the continuum  (Fig.~\ref{fig:l1536zoom}) 
and the \ceo\ integrated map (Fig.~\ref{fig:mom0maps}).  The dominant motion at 
low velocity offsets ($\Delta \upsilon = \pm 2$ \kms) is infall, while a flatter 
slope with a power-law close to $r^{-0.5}$ best describes the high velocity 
components.  If the peak positions were not shifted, only the blue-shifted peaks 
follow the $r^{-0.5}$ profile starting from 70--80 AU.  The flat velocity 
profile is based on 4 points and it does not show a very clear Keplerian disk 
profile such as seen toward TMC1A.  From the high velocity peaks, the velocity 
profile seems to indicate a break at $\sim$80 AU.


\subsection{Summary and stellar masses}

\begin{table}
\centering
 \caption{Derived disk sizes and stellar masses from the break radius in the 
$uv$-space PV diagrams corrected for inclination.}
 \label{tbl:starmass}
 \begin{tabular}{c c c c c}\toprule \hline
  & TMC1A & TMC1 & TMR1 & L1536  \\
  \hline
  $R_{\rm K}$ [AU]          & 80--100  & 100 & $<50$ & 80 \\
  $M_{\star}$ [$M_{\odot}$] &  0.53$^{+0.2}_{-0.1}$  & $0.54^{+0.2}_{-0.1}$ & 
... & 0.7--0.8   \\
  $i$ [$^{\circ}$]          & 55 & 55 & 15 & 65 \\
  \bottomrule
 \end{tabular}

\end{table}

Three out of the four Class I sources show clear indications of embedded RSDs.  
From the turn-over radius (from $r^{-1}$ to $r^{-0.5}$), the radii of 
the Keplerian disks must be between 100--150 AU in TMC1, 80 AU toward L1536 and 
100 AU toward TMC1A.  Considering that the turnover radius is the outer radius 
of the embedded Keplerian disk, the stellar masses are derived from that radius 
and tabulated in Table~\ref{tbl:starmass}.  In the case of L1536, we chose the 
red-shifted peak ($\sim 3$ \kms) at 70 AU as the turnover radius to determine 
the combined stellar masses.  Thus, by obtaining the masses of the different 
components (envelope, disk, and star), it is now possible to place the sources 
in an evolutionary stage for comparison with theoretical models.  The three 
sources are indeed in Stage I of the embedded phase, where the star has accreted 
most of its final mass.

The $uv$-space PV diagram assumes that the disk rotation is perpendicular to the 
outflow direction as inferred from large \mco\ maps, foreground clouds do not 
contaminate the molecular lines, and the source systemic velocity is well 
determined.  Small systemic velocity shifts can indeed shift both 
the red-shifted and blue-shifted velocity profiles from $r^{-0.5}$ to a steeper 
slope.  However, for the firm detection of an embedded RSD toward TMC1A, this 
does not significantly affect the break radius.  Further observations at higher 
resolution and sensitivity are certainly needed to confirm the extent of the 
Keplerian disks toward all of these sources.  However, we argue that even with 
the current data it is possible to differentiate the RSD from the infalling 
rotating envelope with spatially and spectrally resolved molecular line 
observations.


\subsection{Disk gas masses}

\begin{table}
 \centering
 \caption{\ceo\ $J=$2--1 integrated intensities within $R_{\rm K}$ box as 
defined in table~\ref{tbl:starmass}.  Disk masses [10$^{-3}$ $M_{\odot}$] 
inferred from the gas lines and continuum. } 
\label{tbl:masses}
 \begin{tabular}{c c c c c}\toprule \hline
  & TMC1A & TMC1 & TMR1 & L1536  \\
  \hline
  $\int S_{\nu} d\upsilon$\tablefootmark{a} [Jy \kms] &
                        1.93  & 0.63 &  0.26 & 0.17 \\
  $M_{\rm disk}$ gas & 75 & 24 & 10 & 6.8 \\
  $M_{\rm disk}$ dust\tablefootmark{b} & 41--49 & 4.6--5.4 & 10--15 & 19--24 \\
  \bottomrule
 \end{tabular}
 \tablefoot{
 \tablefoottext{a}{Typical errors are 0.03--0.04 Jy \kms.}
 \tablefoottext{b}{Disk masses from the range of dust masses obtained with
the different methods in Sections~\ref{sec:diskdust} 
and~\ref{sec:diskparams} with a gas-to-dust ratio of 100.}}
\end{table}

After determining the extent of the embedded Keplerian disks, their gas masses 
can be estimated from the \ceo\ integrated intensities.  The $J$=2--1 line is 
expected to be thermalized at the typical densities ($n_{\rm H_2} > 10^7$ 
cm$^{-3}$) of the inner envelope due to the low critical density.  The gas mass 
assuming no significant freeze-out is then given by the following equation 
\citep{scoville86, hogerheijde98, momose98}:
\begin{equation}
M_{\rm gas} = 5.45 \times 10^{-4} \frac{T_{\rm ex} + 0.93}{\exp \left ( 
-E_{\rm u} / k T_{\rm ex} \right )} \frac{\tau}{1 - \exp^{-\tau}} \int 
S_{\upsilon} d\upsilon  \  M_{\odot},
\end{equation}
where $h$ and $k$ are natural constants, $T_{\rm ex}$ is the excitation 
temperature, $\tau$ is the line optical depth, $E_{\rm u}$ is the upper energy 
level in K and $\int S_{\nu} d\upsilon$ is the integrated flux densities in Jy 
km s$^{-1}$ using a distance of 140 pc.  The gas mass estimates of 
the disks inferred by integrating over a region similar to the extent of $R_{\rm 
K}$ in Table~\ref{tbl:starmass} are listed in Table~\ref{tbl:masses} with $\tau 
= 0.5$ and $T_{\rm ex} = 40$ K.  The adopted excitation temperature comes from 
the expected \ceo\ rotational temperature within a 1$\arcsec$ beam if a disk 
dominates such emission \citep{harsono13a}.  From the observed $\frac{^{13}{\rm 
CO}}{{\rm C}^{18}{\rm O}}$ flux ratios, the $\tau_{\rm ^{13}CO}$ is estimated 
to be $\leq 4.0$ with isotopic ratios of $^{12}$C/$^{13}$C = 70 and 
$^{16}$O/$^{18}$O = 550 \citep{wilson94}.  This implies that the \ceo\ is 
almost optically thin ($\tau_{\rm C^{18}O} \leq 0.5$) and only a small 
correction is needed under the assumption that \tco\ and \ceo\ trace the same 
structure.  Uncertainties due to the combined $T_{\rm ex}$ and $\tau$ result in 
a factor of $\le 2$ uncertainty in the derived disk masses.  The gas disk masses 
toward TMC1A, TMC1 and TMR1 are at least a factor of 2 higher than the mass 
derived from the continuum data (assuming gas to dust ratio of 100).  Since the 
velocity gradient is prominent in the moment one maps, there must be enough 
\ceo\ in the gas phase within the RSD in order to significantly contribute to 
the lines.  This is consistent with the evolutionary models of \citet{visser09} 
and \citet{harsono13a} where only a low fraction of CO is frozen out within 
the embedded disks unless $R_{\rm K} > 100$ AU due to the higher luminosities.  
Only toward L1536, which is the most evolved source in our sample, is the gas 
mass a factor of 5 lower than the mass derived from the continuum.  This would 
be consistent with CO starting to freeze-out within the disk near the end of the 
main accretion phase \citep{visser09}.  Thus, the small differences between gas 
and dust masses indicate relatively low CO freeze-out in embedded disks.


\section{Discussion}\label{sec:dis}

\subsection{Disk structure comparison: dust versus gas}

In this paper, we have determined the sizes of the disk-like structures around 
Class I embedded YSOs using both the continuum and the gas lines.  The 
continuum analysis focuses on the extent of continuum emission excluding the 
large-scale emission, while the gas line analysis focuses on placing 
constraint on the size of the RSD.  The continuum analysis utilizes both 
intensity based disk models and a simple disk structure (power-law disk model). 
 The intensity based disk models give a large spread of disk radii toward all 
sources.  On the other other hand, the best-fit radii of power-law disks are 
larger than the Keplerian disk toward TMR1 and L1536, while the values are
comparable toward TMC1A and TMC1.  Not all of the disk structure derived 
from continuum observations may be associated to an RSD.

The main caveat in deriving the flattened disk structure from continuum 
visibilities is the estimate of the large-scale envelope emission.  It has been 
previously shown that the large-scale envelope can deviate from spherical 
symmetry \citep[e.g.,][]{looney07, tobin10a}.  Disk structures may change if 
one adopts a 2D flattened envelope structure due to the mass distribution at 
50--300 AU scale.  However, in this paper, we focused on the size of the 
flattened structure where any deviation from the spherically symmetry model 
gives an estimate of its size.  For the purpose of this paper, a spherical 
envelope model is used to estimate the large-scale envelope contribution since 
it is simple, fits the observed visibilities at short baselines, and does not 
require additional parameters.


\subsubsection{Comparison to disk sizes and masses from SED modelling}

A number of 2D envelope and disk models have been published previously 
constrained solely using continuum data.  More recently, \citet{eisner12} used 
the combination of high resolution near-IR images, the SED and millimeter 
continuum visibilities to constrain the structures around TMR1, TMC1A and L1536. 
 The size of his disk is defined by the centrifugal radius, which is the 
radius at which the material distribution becomes more flattened 
\citep{ulrich76}.  His inferred sizes are consistent with the extent of RSDs in 
our sample (TMR1: 30--450 AU; L1536: 30--100 AU; TMC1A: 100 AU), but our 
higher resolution data narrow this range.  For the case of TMR1, we can 
definitely rule out any disk sizes $>100$ AU.

Others have used a similar definition of the centrifugal radius using the 
envelope structure given by \citet{ulrich76} with and without disk component 
\citep[e.g.,][]{eisner05, gramajo07, furlan08}.  Thus, the value of $R_{\rm c}$ 
gives an approximate outer radius of the RSD.  These models are then fitted to 
the SED and high resolution near-IR images without interferometric data.  In 
general, the values of $R_{\rm c}$ derived from such models are lower (at 
least by factor of 2) than the extent of the RSDs as indicated by our molecular 
line observations. By trying to fit the models to the near-IR images which are 
dominated by the hot dust within the disk, they put more weight to the inner 
region.  Thus, we argue that spatially resolved millimeter data are necessary in 
order to place better constraints on the compact flattened structure.

The disk masses derived in our work from both 50 k$\lambda$ flux and $uv$ 
modelling are consistent with each other.  These values are typically higher by 
a factor of 2 and a factor of 8 for TMC1A\footnote{We note that our observed 
visibilities (Fig.~\ref{fig:uvamp}) are a factor of 3 higher than that reported 
by \citet{eisner12} for TMC1A at 50 k$\lambda$.} than the disk masses reported 
by \citet{eisner12} and \citet[][only toward TMR1]{gramajo10}.  The discrepancy 
may be due to the difference in the combination of the assumed inclination, the 
adopted envelope structure, and definition of the flattened structure.


\subsection{Constraints on disk formation}


\begin{table*}
 \centering
 \caption{Properties of observed embedded RSDs.}
 \label{tbl:obsrsds}
 \begin{tabular}{c c c c c c c c c c }\toprule \hline
 Source & $M_{\star}$ & $M_{\rm disk}$ & $R_{\rm K}$\tablefootmark{a} & 
$M_{\star}/M_{\rm tot}$\tablefootmark{b} & $ M_{\rm disk} / M_{\star} $ & 
$\Theta_{\rm disk}$\tablefootmark{c} & $\Theta_{\rm rotation}$\tablefootmark{d} 
& $\lambda_{\rm eff}$\tablefootmark{e} & References \\
 & [$M_{\odot}$] & [$M_{\odot}$] & [AU] & & & [deg] & [deg] &  & \\
 \hline
  \multicolumn{9}{c}{Class 0} \\
  NGC1333 IRAS4A2   & 0.08 & 0.25 & 310 & 0.02 & 3.1 & 108.9 & ... & ... & 
1,2,3 \\
    L1527           & 0.19 & 0.029--0.075 & 90 & 0.2 & 0.13--0.34 & 0 & -177 & 
... & 1,2,4,5 \\
  VLA1623           & 0.20 & 0.02 & 150 & 0.4--0.6  & ... & -145 & ... & ... & 
6,7  \\
  \multicolumn{9}{c}{Class I} \\
  R CrA IRS7B          & 1.7 & 0.024 & 50 & 0.43  & 0.01 & -65 & ... & ... 
& 8   \\
  L1551 NE          & 0.8 & 0.026 & 300 & 0.65  & 0.032 & 167 & ...  & ... & 9 
 \\
  L1489-IRS         & 1.3 & 0.004 & 200 & 0.83--0.93 & 0.0030 & -120 & 110 
& ... & 1,2,10 \\
  IRS43             & 1.9 & 0.004 & 190 & 0.89  & 0.002 & 107 & ...  & ...  
& 1, 12 \\
  IRS63             & 0.8 & 0.099 & 165 & 0.83  & 0.12 & -10 & ...  & ...  
& 11, 12 \\
  Elias29           & 2.5 & 0.011  & 200 & 0.98  & $< 0.003$ & ...  & ...  & 
...  & 11 \\
  TMC1A             & 0.53 & 0.045--0.075  &  100 & 0.75--0.78  & 0.08--0.14 & 
-115  & 25 & $>10$  & 13  \\
  TMC1              & 0.54 & 0.005--0.024  &  100 & 0.76--0.79  & 0.01--0.06 & 
-90 &  35 & 4 & 13  \\
  TMR1              & 0.7 & 0.01--0.015  &  $<50$ & 0.72\tablefootmark{f} &  
0.02--0.03  & ...  & ...  & ...  & 13  \\
  L1536\tablefootmark{g} & 0.4 & 0.007--0.024 & 80 & 0.95--0.97 &  0.02--0.06  
& -95 & -7 & ...  & 13  \\
  \bottomrule
 \end{tabular}
 \tablefoot{
 \tablefoottext{a}{Outer radius of the Keplerian disk.}
 \tablefoottext{b}{$M_{\rm tot} = M_{\star} + M_{\rm disk} + M_{\rm env}$. }
 \tablefoottext{c}{Direction of the disk rotation from blue to red.}
 \tablefoottext{d}{Direction of the envelope rotation associated to or near 
the sources from \citet{caselli02} except for L1527 whose value is from 
\citet{goodman93}.}
 \tablefoottext{e}{Effective mass-to-flux ratio scaled from the average value 
in \citet{troland08} with magnetic field strengths from \citet{crutcher10}.}
\tablefoottext{f}{Stellar mass adopted from \citet{hogerheijde98}, which 
assumes that all of the luminosity is due the star.}
\tablefoottext{g}{The stellar mass is sum of the two stars in the binary. } }
\tablebib{
(1)~\citealt{prosac09}; (2)~\citealt{kristensen12}; (3)~\citealt{choi10}; 
(4)~\citealt{tobin12}; (5)~\citealt{tobin13}; (6)~\citealt{murillo13a}; 
(7)~\citealt{murillo13b}; (8) Lindberg et al. subm.; (9)~\citealt{takakuwa12}; 
(10)~\citealt{brinch07}; (11)~\citealt{lommen08}; (12)~\citealt{brinch13}; (13) 
This work}
\end{table*}

\begin{figure}
 \centering
 \includegraphics{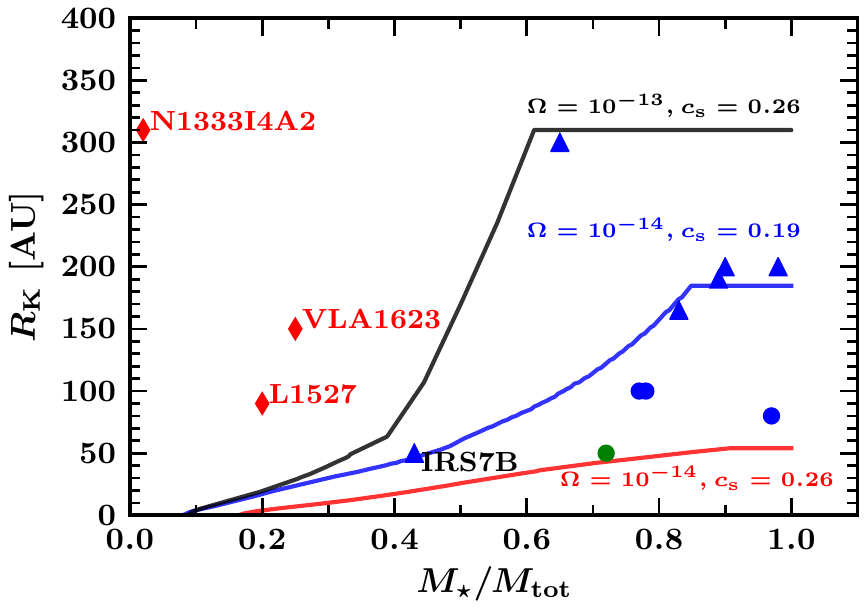}
 \caption{Comparison of the observed RSDs (symbols) and semi-analytical 
models (lines) from \citet{harsono13a}.  The red diamonds indicate the Class 0 
sources, blue circles show the Class I sources we have analyzed, and blue 
triangles show the Class I sources whose parameters were taken from the 
literatures.  TMR1 is indicated by the green circle since the stellar mass 
is derived from the bolometric luminosity.  Different lines show the evolution 
of RSDs with different initial conditions ($c_{\rm s} = 0.26$ \kms\ (black and 
red) and $0.19$ \kms\ (blue); $\Omega = 10^{-13}$ Hz (black) and $10^{-14}$ Hz 
(red and blue)), keeping the initial core mass fixed at 1 M$_\odot$. }
\label{fig:comprdisk}
\end{figure}


A number of RSDs toward other embedded YSOs have been reported in the 
literature, which are summarized in Table~\ref{tbl:obsrsds}.  The values are 
obtained from analysis that included  interferometric molecular line 
observations.  Keplerian radii, $R_{\rm K}$, between 80 and 300 AU are found 
toward these sources with disk masses in the range 0.004--0.055 $M_{\odot}$ 
around 0.2--2.5 $M_{\odot}$ stars.  Most of the observed RSDs are found toward 
Stage I sources ($M_{\star} > M_{\rm env}$), while RSDs are claimed toward three 
Stage 0/I sources.  From the sample, large ($\ge 100$ AU) RSDs are generally 
found toward sources where $M_{\star}/M_{\rm tot} > 0.65$ with $M_{\rm tot} = 
M_{\star} + M_{\rm env} + M_{\rm disk}$.

How do these disks compare to disk formation models? Semi-analytical disk 
formation models by \citet{visser09}, which are based on \citet{tsc84} and 
\citet{cm81}, predict that $>100$ AU RSDs are expected by the end of the 
accretion phase \citep[$M_{\rm env} < M_{\rm disk}$, see also][]{young05}.  
The disk sizes depend on the initial envelope temperature or sound speed and 
the rotation rate.  The evolution of $R_{\rm K}$ within these models is shown in
Fig.~\ref{fig:comprdisk} for the same initial conditions as \citet{harsono13a} 
with comparable initial rotation rates to those measured by \citet{goodman93}.  
Remarkably, the observed RSDs in Stage I fall within the expected sizes from 
these semi-analytical models.  On the other hand, most of the Class 0 disks 
fall outside of these models.  From this comparison, the observed RSDs in Stage 
I can be well represented by models with $\Omega = 10^{-14}$ Hz and $c_{\rm s} 
=$ 0.19--0.26 \kms.

In semi-analytical models, the disk forms as a consequence of angular momentum 
conservation.  However, recent numerical simulations of collapse and disk 
formation with the presence of magnetic fields do not always form a large RSD 
($> 100$ AU).  \citet{galli06} showed that an RSD does not form in the ideal 
magnetohydrodynamics (MHD) case due to efficient magnetic breaking \citep[e.g., 
][]{galli03, mellon08}.  The problem can be alleviated if the magnetic field is 
weak or with the inclusion of non-ideal MHD, although the results with non-ideal 
MHD are not entirely clear \citep{mellon08, mellon09, dapp10, krasnopolsky11, 
zyli11, braiding12}.  Once the RSD forms, it can easily evolve to $>400$ AU 
\citep{dapp12, vorobyov10b, joos12}.  The question centers on whether RSDs are 
already present in the Class 0 phase.  This depends on the strength and 
orientation of the magnetic field, as well as the initial rotation rate of the 
envelope.


\subsubsection{On the magnetic field}

The dynamical importance of the magnetic field can be numerically determined 
through the value of the effective mass-to-flux ratio, $\lambda_{\rm eff}$, 
which is the ratio of the gravitational and magnetic energies within the cloud. 
 A value of 1 means that the cloud is strongly magnetized while value of $>10$ 
characterizes weakly magnetized clouds.  \citet{troland08} presented a recent 
survey of magnetic field strengths toward molecular cloud cores.  They find 
an average $\lambda_{\rm eff} \sim 2-3$ or an average magnetic field strength 
of 16 $\mu$G, which argues for the importance of magnetic field during the 
dynamical evolution of pre- and protostellar cores \citep[see 
also][]{crutcher12}.

The magnetic field strength within Taurus is found to vary toward different 
cores.  For example, TMC1A is within the L1534 core whose line of sight 
magnetic field is weak \citep[$<1.2 \ \mu$G,][]{crutcher10}, while the TMC1 dark 
cloud has considerable magnetic field strength of the order of $9.1 \pm 2.2\ 
\mu$G.  Since $\lambda$ is inversely proportional to the magnetic field 
strength, the $\lambda_{\rm eff}$ toward these two sources are $> 10$ and $\sim 
4$ around TMC1A and TMC1, respectively.  It is interesting to note that a large 
and massive RSD ($R = 10 0$ AU and $M_{\rm disk}/M_{\star} \sim 0.1$) is found 
toward TMC1A, while the RSDs toward both TMC1 and TMR1, which are located 
closer to the TMC1 dark cloud, are small and less massive.  Considering that 
$M_{\rm  disk}/M_{\rm env}$ is lower toward TMC1 and TMR1 than toward TMC1A, 
this also illustrates that the magnetic field may play an important role in 
determining the disk properties.

How likely are these disks formed out of weakly magnetized cores? Molecular 
cloud cores are found to have lower than average magnetic field strength 
\citep[$\lambda_{\rm eff} > 3$,][]{crutcher10}, than found on cloud scales.  On 
the other hand, the fraction of cores with $\lambda_{\rm eff} > 5$ is low.  
Consequently, it is unlikely that TMC1A and other observed RSDs in Stage I all 
formed out of weakly magnetized cores.  Thus, simulations need to be able to 
form RSDs out of a moderately magnetized core that evolve up to sizes of 
100--200 AU by Stage I.  In addition, we can argue that the magnetic breaking 
efficiency drops as the envelope dissipates, which allows the rapid growth of 
RSDs near the end of the main accretion phase as shown in 
Fig~\ref{fig:comprdisk}.  On the other hand, there may be a low fraction of 
cores which are weakly magnetized, which facilitate formation of large disks as 
early as Stage 0 phase \citep{krumholz13, murillo13b}.


\subsubsection{On the large-scale rotation}

Another important parameter is the large-scale rotation around these sources, 
which is a parameter used in both simulations and semi-analytical models.  
Generally, a cloud rotation rate between $10^{-14}$ Hz to a few $10^{-13}$ Hz 
is used in numerical simulations \citep[e.g.,][]{yorke99, zyli11}.  The average 
rotation rate measured by \citet{goodman93} is $\sim 10^{-14}$ Hz, which is 
consistent with the initial rotation rate of the semi-analytical model that best
describes a large fraction of the observed RSDs.  The time at which Stage I 
starts within that model is $3\times10^5$ years.  For comparison, around the 
same computational time, \citet{yorke99} form a $>500$ AU RSD for their low-mass 
case (J).  Similarly, \citet{vorobyov10b} also found $>200$ AU RSDs close to 
the start of the Class II phase.  We find no evidence of such large RSDs at 
similar evolutionary stages in our observations.

Another aspect of the large-scale rotation that is worth investigating is its 
direction with respect to that of the disk rotation.  Models assume that the 
disk rotation direction is the same as the rotation of its core.  To compare 
the directions, we have used the velocity gradients reported by 
\citet{caselli02} and \citet{goodman93} toward the dark cores in Taurus.  
Velocity gradients are detected toward L1534 (TMC1A), L1536 and TMC-1C, which 
is located to the North East of TMR1 and TMC1.  Table~\ref{tbl:obsrsds} lists 
the velocity gradients of the RSDs toward the embedded objects and their 
envelope/core rotation\footnote{The direction is in degrees East of North and   
increasing $\upsilon_{\rm lsr}$ (blue to red) to be consistent with 
\citet{caselli02} and \citet{goodman93}.}.  It is interesting to note that the 
large-scale rotation detected around or nearby the RSDs have a different 
velocity gradient direction.  Similar misalignments in rotation have been 
pointed out by \citet{takakuwa12} toward L1551 NE and by \citet{tobin13} for 
L1527.  Although the positions of the detected molecular emission in 
\citet{caselli02} do not necessarily coincide with the embedded objects due to 
chemical effects, the systematic velocities of the cores are similar.  
Misalignment at 1000 AU scales was also concluded by \citet{brinch07} toward 
L1489 where they found that the Keplerian disk and the rotating envelope are at 
an angle of 30$^{\circ}$ with respect to each other.  If these objects formed 
out of these cores in the distant past, then either the collapse process 
changed the angular momentum of the envelope and, consequently, the disk 
rotation direction or it affects the angular momentum distribution of nearby 
cores through feedback.  Interestingly, the former may suggest the importance of 
the non-ideal MHD effect called the Hall effect during disk formation, which can 
produce counter-rotating disks \citep{zyli11, braiding12}.

Finally, the simulations and models above require the initial core to be 
rotating in order to form RSDs.  However, by analyzing the angular momentum 
distribution of magnetized cores, \citet{dzib10} suggested that the NH$_3$ and 
N$_2$H$^{+}$ measurements may overestimate the angular momentum by a factor of 
$\sim 8-10$.  This poses further problems to disk formation in previous 
numerical studies of collapse and disk formation.  On the other hand, the 
inclusion of the Hall effect does not require a rotating envelope core to form a
rotating structure \citep[e.g.,][]{krasnopolsky11,zyli11}.  Further studies are 
required to compare the expected observables from rotating and non-rotating 
models.


\subsection{Disk stability}

Self gravitating disks can be important during the early stages of star 
formation in regulating the accretion process onto the protostar, which can be 
variable \citep[e.g.,][]{vorobyov10b}.  \citet{dunham12} show that such episodic 
accretion events can reproduce the observed population of YSOs.  A 
self-gravitating disk is expected when $M_{\rm disk}/ M_{\rm star} \ge 0.1 $ 
\citep[e.g.,][]{lr04, boley06, cossins09}.  However, most of the observed 
embedded RSDs around Class I sources have disk masses such that $M_{\rm disk}/ 
M_{\rm star} \le 0.1$.  It is unlikely for such disks to become 
self-gravitating, but they may have been in the past.  Interestingly, a few 
embedded disk seem to be self-gravitating since their $M_{\rm disk} / 
M_{\rm star} \ge 0.1$ such as toward TMC1A, IRS63, L1527 and NGC1333 IRAS4A2 
(See Table~\ref{tbl:obsrsds}).

The low fraction of large and massive disks during the embedded phase suggests 
that disk instabilities may have taken place in the past and limited their 
current observed numbers (i.e. low fluxes at long baselines).  In this scenario, 
the embedded disks will have a faster evolution and, thus, spend most of their 
lifetime with small radii.  Alternatively, disks experiencing high infall rates 
from envelope to the disk as expected during the embedded phase may also have 
higher accretion rates \citep{hal11}.  Such a scenario argues for disk 
instabilities to be present even for $M_{\rm disk}/M_{\star} < 0.1$.  In both 
cases, the dominant motion of the compact flattened structure during the 
embedded phase is infall rather than rotation.  There are indeed sources where 
higher sensitivity and spatial resolution observations such as with Atacama 
Large Millimeter/submillimeter Array (ALMA) can detect features associated with 
disk instabilities \citep[e.g.,][]{cossins10,forgan12,douglas13}.


\section{Conclusions}\label{sec:sum}

We present spatially and spectrally resolved observations down to a radius of 
56 AU around four Class I YSOs in Taurus.  The \ceo\ and \tco\ 2--1 lines are 
used to differentiate the infalling rotating envelope from the rotationally 
supported disk.  Analysis of the dust and gas lines were performed directly in 
$uv$ space to avoid any artefacts introduced during the inversion process.  
The main results of this paper are:

\begin{itemize}
\item Dust disk sizes and masses can be derived from the continuum by using 
    a power-law spherical envelope model to account for the large-scale 
    envelope emission.  Intensity based disk models (Uniform and Gaussian) give 
    similar disk sizes to each other which are lower by typically 
    25\%-90\% than realistic disk models (power-law disk).  On the other hand, 
    they give similar disk masses to more realistic models, which are 
    consistent with the disk mass derived from the 50 k$\lambda$ flux point.  
    Inclusion of the envelope in the analysis is important to obtain 
    reliable disk masses.  The observationally derived masses indicate that 
    there is still a significant envelope present toward three of the four 
    sources: TM1A, TMC1, and TMR1.

\item Three of the four sources (TMC1, L1536, TMC1A) host embedded rotationally 
    supported disks (RSDs) derived from line data.  By fitting velocity 
    profiles ($\upsilon \propto r^{-0.5}$ or $\propto r^{-1.0}$) to the red- 
    and blue-shifted peaks, the RSDs are found to have outer radii of 
    $\sim 80$--100 AU.  In addition, the large-scale structure ($>100$ AU) is 
    dominated by the infalling rotating envelope ($\upsilon \propto r^{-1}$). 
    The derived stellar masses of these sources are of order 0.4--0.8 
    $M_{\odot}$, consistent with previous values.  Consequently, these objects 
    are indeed Stage I young stellar objects with ${M_{\star}}/({M_{\rm 
    env}+M_{\rm disk}+M_{\star})} > 0.7$.

\item Disk radii derived from the power-law dust disk models toward TMC1 
    and TMC1A are consistent with sizes of their RSDs.  However, the dust disk 
    radii toward TMR1 and L1536 are not the same as the extent of their RSDs.  
    Thus, we emphasize that spatially and spectrally resolved gas lines 
    observations are required to study the nature of flattened structures 
    toward embedded YSOs.

\item Semi-analytical models with $\Omega = 10^{-14}$ Hz and $c_{s} =$ 0.19 
    -- 0.26 \kms  describe most of the observed RSDs in Stage I.  The observed 
    RSDs argue for inefficient magnetic breaking near the end of the main 
    accretion phase ($M_{\star}/M_{\rm tot} > 0.65$).  More theoretical studies 
    are needed to understand how and when RSDs form under such conditions.

\item Comparison between disk masses derived from the continuum and \ceo\ 
    integrated line intensities (Table~\ref{tbl:masses}) suggests that 
    relatively little of CO is frozen out within the embedded disk.

\end{itemize}

The current constraints on disk formation rely on a small number of observed 
RSDs around Class I sources.  Certainly one needs to constrain the size of RSDs 
near the end of the main accretion phase to understand how late Class II disks 
form.  On the other hand, the observed RSDs in the Class I phase do not answer 
the question when and how RSDs form.  Such constraints require the detection and 
characterization of RSDs during Class 0 phase, which are currently lacking.  
From the sample of sources in \citet{yen13} and our results, it is clear that 
the infalling rotating flattened structure is present at $>$100 AU.  Thus, 
extending molecular line observations down to $<50$ AU radius with ALMA toward 
Class 0 YSOs will be crucial for differentiating between the different scenarios 
for disk formation.


\section*{Acknowledgements}

We thank the anonymous referee for the constructive comments.  We are grateful 
to the IRAM staff, in particular to Chin-Shin Chang, Tessel van der Laan and 
J{\'e}r{\^o}me Pety, for their help with the observations, reduction and 
modelling of the data.  We thank Umut Y{\i}ld{\i}z for reducing the single-dish 
data and making them available.  This work is supported by the Netherlands 
Research School for Astronomy (NOVA) and by the Space Research Organization 
Netherlands (SRON).  Astrochemistry in Leiden is supported by the Netherlands 
Research School for Astronomy (NOVA), by a Spinoza grant and grant 614.001.008 
from the Netherlands Organisation for Scientific Research (NWO), and by the 
European Community's Seventh Framework Programme FP7/2007–2013 under grant 
agreement 238258 (LASSIE).  This research  was supported by a Lundbeck 
Foundation Group Leader Fellowship to Jes  J{\o}rgensen.  Research at Centre 
for Star and Planet Formation is funded by  the Danish National Research 
Foundation and the University of Copenhagen's  programme of excellence.  This 
research used the facilities of the Canadian  Astronomy Data Centre operated by 
the National Research Council of Canada with  the support of the Canadian Space 
Agency.

\bibliographystyle{aa}
\bibliography{../../biblio}


\appendix 
\Online

\section{Observational data}\label{app:data}

\begin{figure}
 \centering
 \includegraphics{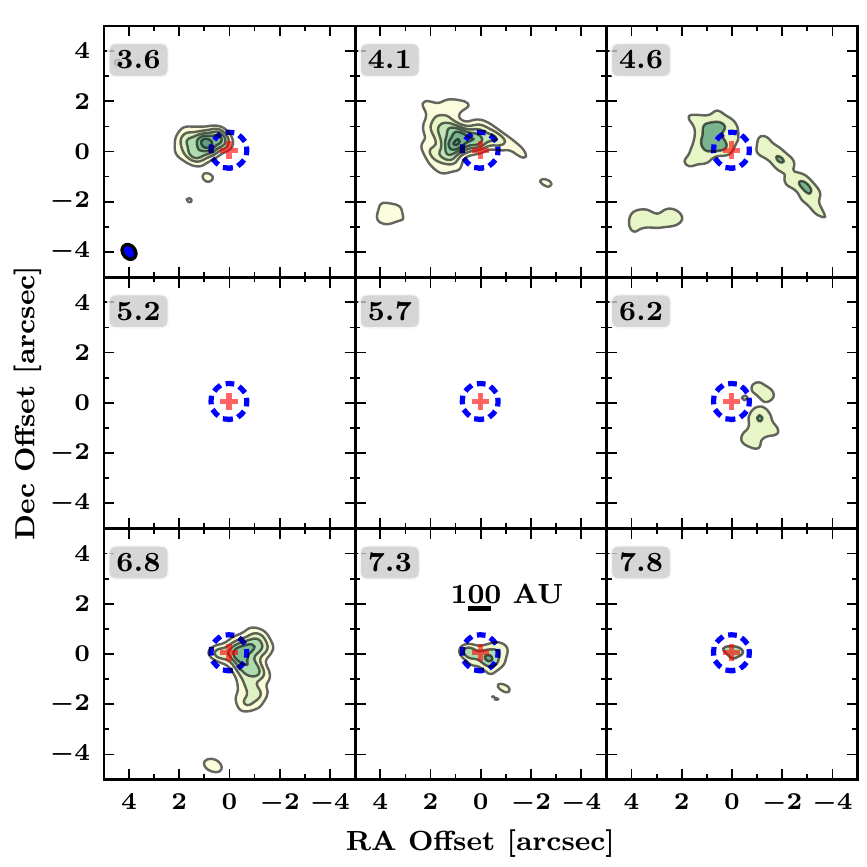}
\caption{\tco\ 2--1 channel maps similar to Fig.~\ref{fig:onemap} toward TMC1.}
\label{fig:tmc1co13}
\end{figure}

\begin{figure}
 \centering
 \includegraphics{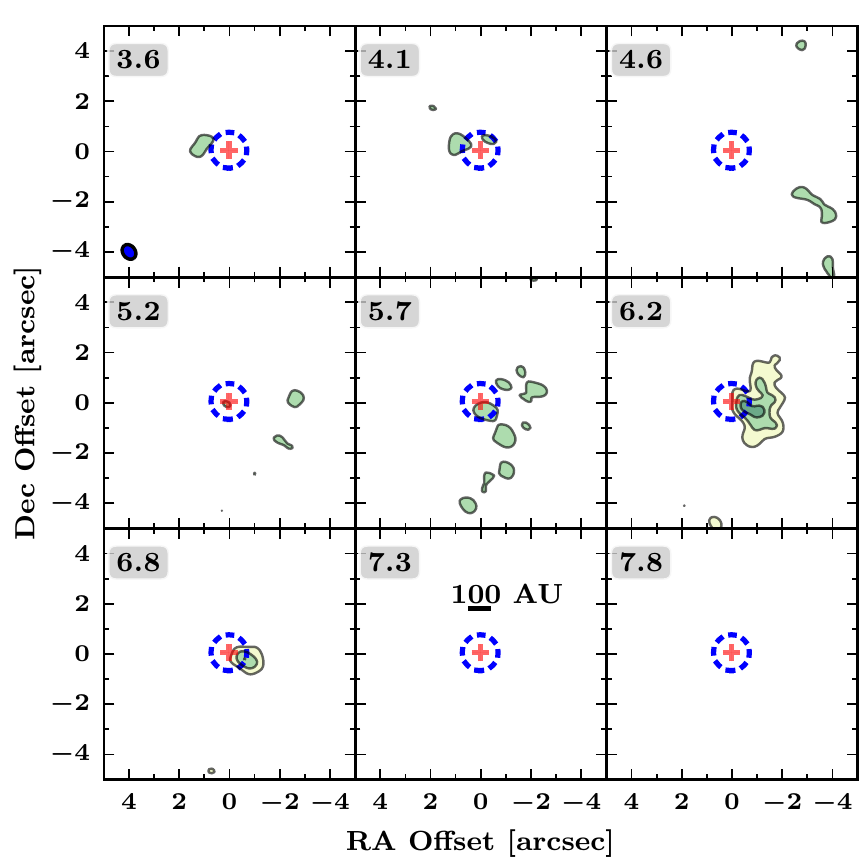}
\caption{\ceo\ 2--1 channel maps similar to Fig.~\ref{fig:onemap} toward TMC1.}
\label{fig:tmc1co18}
\end{figure}

\begin{figure}
 \centering
 \includegraphics{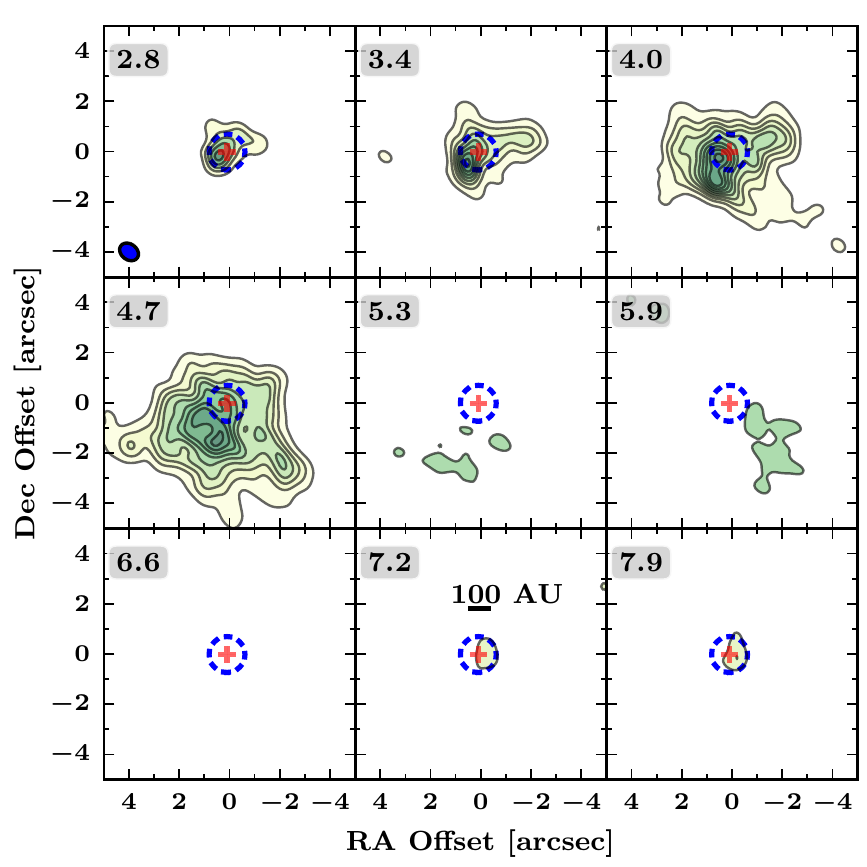}
\caption{\tco\ 2--1 channel maps similar to Fig.~\ref{fig:onemap} toward TMR1. }
\label{fig:tmr1co13}
\end{figure}

\begin{figure}
 \centering
 \includegraphics{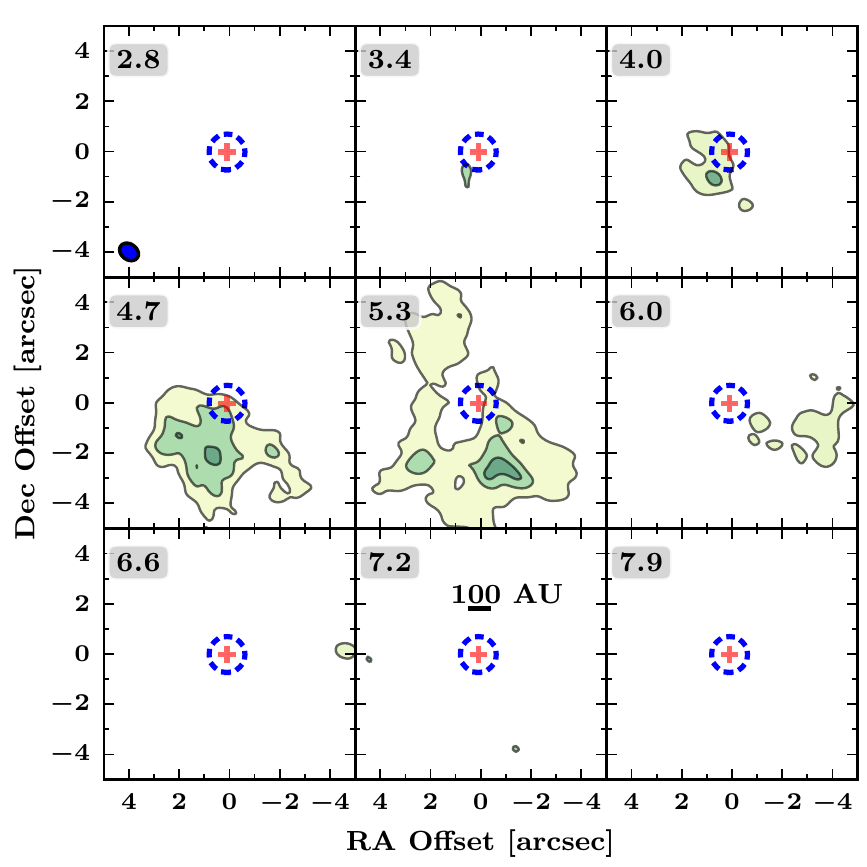}
\caption{\ceo\ 2--1 channel maps similar to Fig.~\ref{fig:onemap} toward TMR1. }
\label{fig:tmr1co18}
\end{figure}

\begin{figure}
 \centering
 \includegraphics{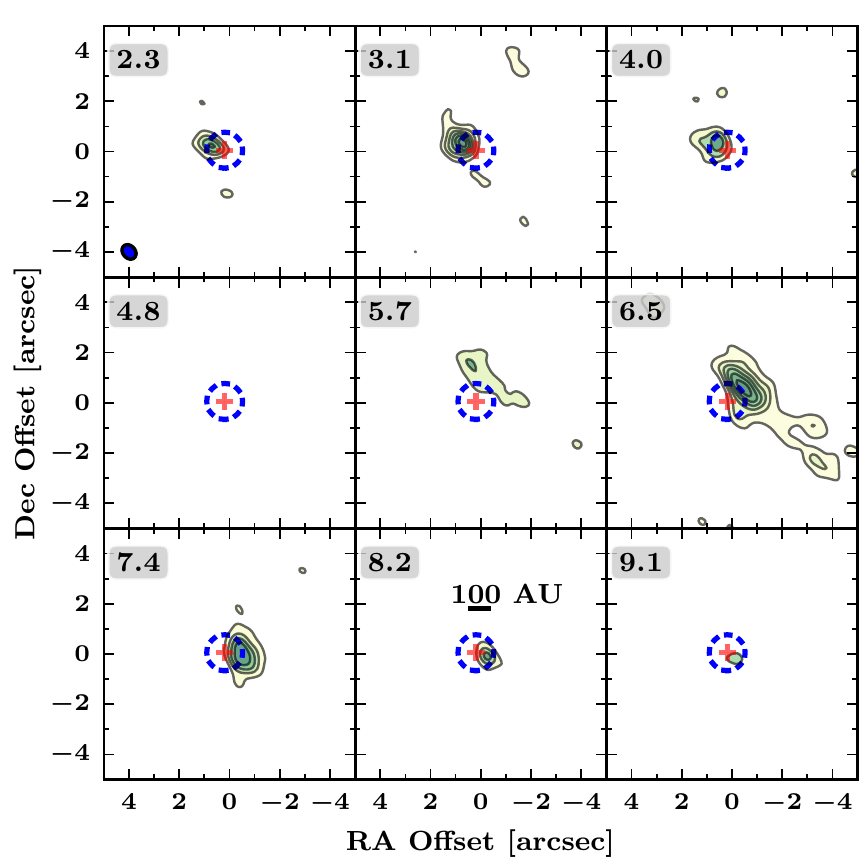}
\caption{\tco\ 2--1 channel maps similar to Fig.~\ref{fig:onemap} toward L1536. 
}
\label{fig:l1536co13}
\end{figure}

\begin{figure}
 \centering
 \includegraphics{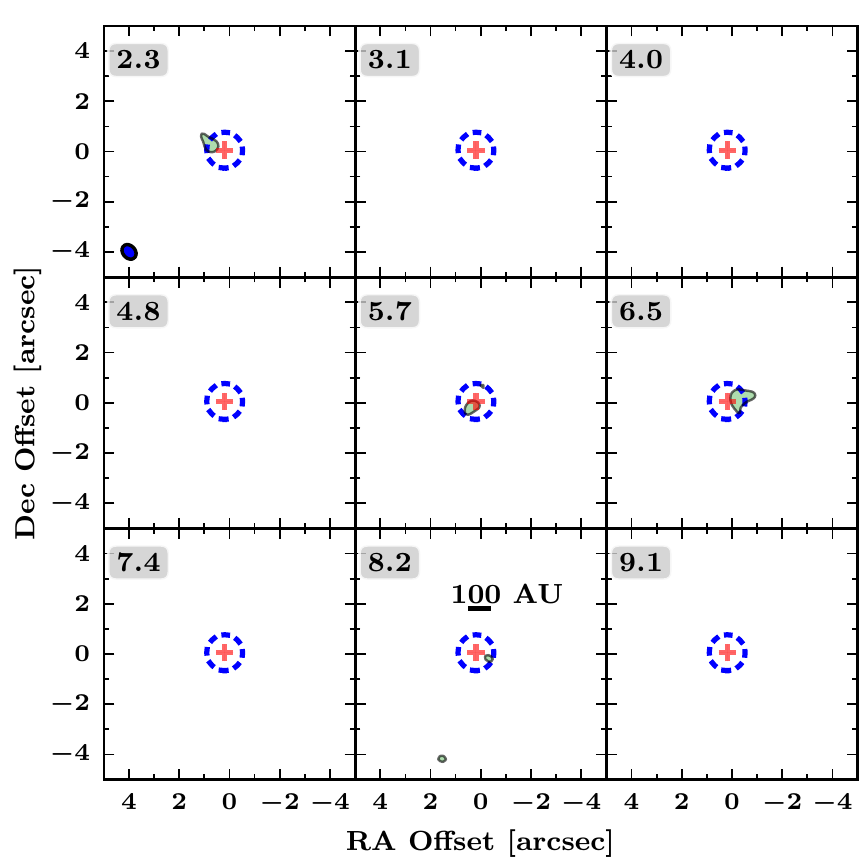}
\caption{\ceo\ 2--1 channel maps similar to Fig.~\ref{fig:onemap} toward L1536. 
}
\label{fig:l1536co18}
\end{figure}

The observational log can be found in Table~\ref{tbl:obsdetails} and the 
channel maps of all the observed lines toward the four sources are shown in 
Figs.~\ref{fig:tmc1co13}--~\ref{fig:l1536co18}.  

\begin{table*}[htbp]
\centering
 \caption{Observational log}
 \label{tbl:obsdetails}
 \begin{tabular}{ c c c c  c }\toprule \hline
 Date & Configuration & No. Antenna & 
Bandpass calibrators (flux) & Gain 
calibrators (flux)  \\
 \hline
 \multicolumn{5}{c}{Track sharing of TMR1 and TMC1A} \\
 \hline
 2012 Mar 3 & B & 6 & 0234+285 (3.07 Jy)  & 0400+258 (0.37 Jy), 0507+179 (2.71 
Jy) \\
  Mar 7 & B & 6 & 3C84 (9.93 Jy)  & 0400+258 (0.35 Jy), 0507+179 (2.54 Jy)  \\
  Mar 10 & B & 6 & 3C84 (9.44 Jy) & 0400+258 (0.33 Jy), 0507+179 (2.35 Jy)  \\
  Mar 12 & B & 5 & 3C84 (8.68 Jy) & 0400+258 (0.29 Jy), 0507+179 (1.88 Jy) \\
  Mar 12 & B & 6 & 3C84 (9.95 Jy) & 0400+258 (0.47 Jy), 0507+179 (3.12 Jy)  \\
  Mar 31 & C & 6 & 2200+420 (7.05 Jy) & 0400+258 (0.28 Jy), 0507+179 (1.79 Jy) 
\\
 \hline
 \multicolumn{5}{c}{Track sharing of TMC1 and L1536} \\
 \hline
 2013 Mar 2 & B & 6 & 3C84 (11.86 Jy)  & 0400+258 (0.29 Jy), 0459+252 (0.35 Jy) 
\\
  Mar  3 & B & 6 & 3C84 (14.29 Jy)  & 0400+258 (0.34 Jy), 0459+252 (0.40 Jy)  \\
  Mar 16 & B & 6 & 3C84 (12.40 Jy) & 0400+258 (0.25 Jy), 
0459+252 (0.21 Jy)   \\
  Mar 19 & B & 6 & 3C84 (10.74 Jy) & 0400+258 (0.26 Jy), 
0459+252 (0.29 Jy) \\
  Mar 22 & B & 6 & 2200+420 (12.18 Jy)  & 0400+258 (0.29 Jy), 
0459+252 (0.28 Jy)  \\
  Mar 27 & B & 6 & 3C84 (12.88 Jy) & 0400+258 (0.28 Jy), 
0459+252 (0.34 Jy)\\   
  Apr  7 & C & 6 & 3C84 (10.32 Jy) & 0400+258 (0.25 Jy),0459+252 (0.32 Jy)  \\
 \bottomrule
 \end{tabular}
\end{table*}

\section{Large-scale structure}

The parameters of the large-scale envelope structure used to estimate the 
continuum emission at short baselines are in Table~\ref{tbl:1dmodels}.

\begin{table}
 \centering
 \caption{1D spherical envelope parameters as published by 
\citet{kristensen12}.}
 \label{tbl:1dmodels}
 \begin{tabular}{cccccc}\toprule \hline
   Source & $p$ & $Y$\tablefootmark{a} & $r_{\rm in}$ & $n_{\rm in}$ & 
$n(1000 {\rm AU})$ 
\\
    &  &  & [AU] & [cm$^{-3}$] & [cm$^{-3}$] \\
   \hline
   TMC1 & 1.1 & 1800 & 3.7 &  $8.5 \times 10^7$ & $1.8\times 10^{5}$ \\
   TMR1 & 1.6 & 900 & 8.8 & $4.1 \times 10^8$ & $2.1\times 10^{5}$  \\
   TMC1A & 1.6 & 900 & 7.7 & $5.2 \times 10^8$ & $2.2\times 10^{5}$ \\
  \bottomrule
   \end{tabular}
   \tablefoot{
   \tablefoottext{a}{$Y = r_{\rm out}/r_{\rm in}$}.
   }
\end{table}


\section{WideX data}\label{app:widex}

\begin{figure*}[htbp]
 \centering
 \includegraphics{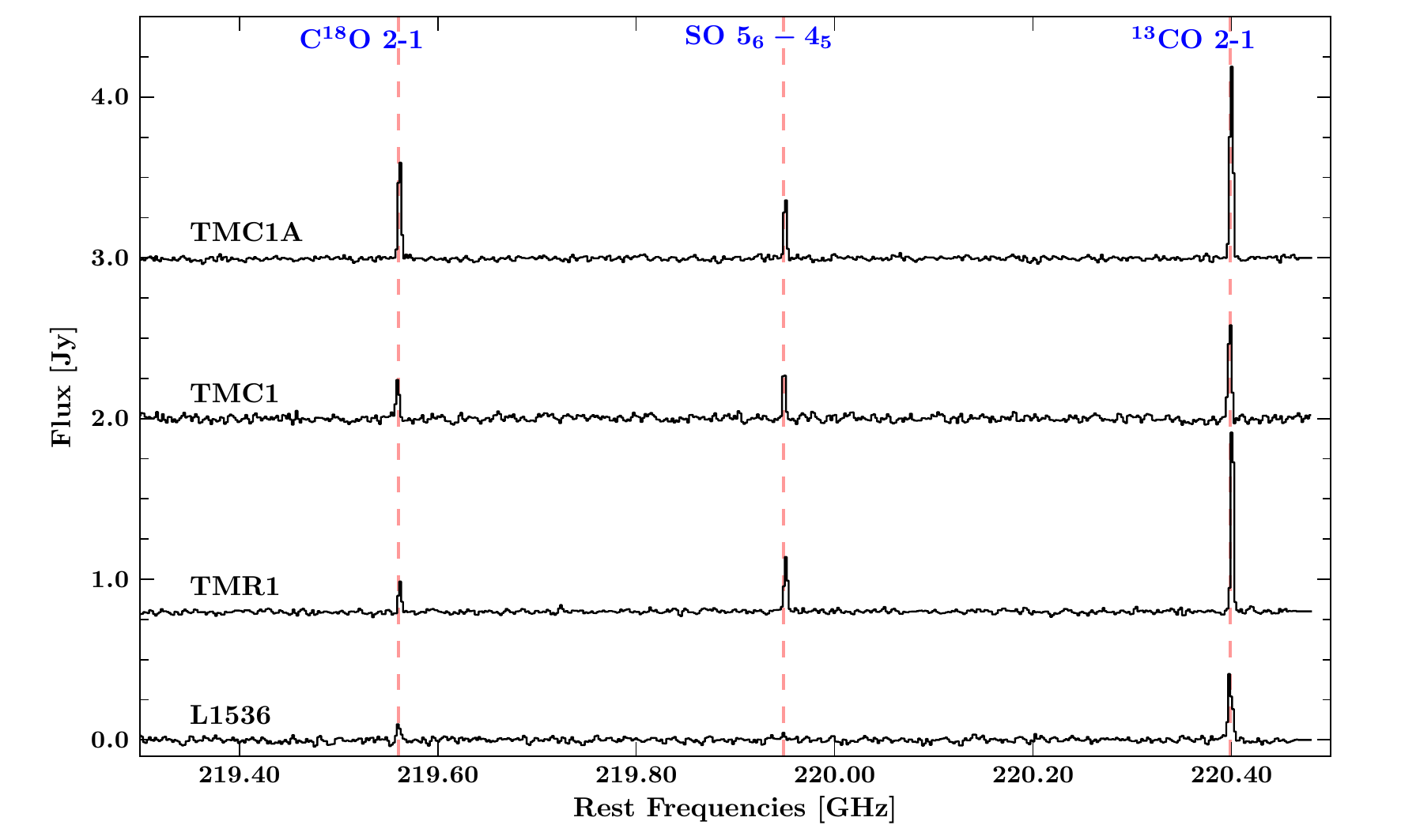}
 \caption{Integrated flux of \ceo\ $J=$2--1, SO $5_6 - 4_5$ and \tco\ $J=$2--1 
within a 3$\arcsec$ box around each source in the WideX spectra.}
  \label{fig:allwidexspecs}
\end{figure*}

\begin{figure*}[htbp]
 \centering
 \includegraphics{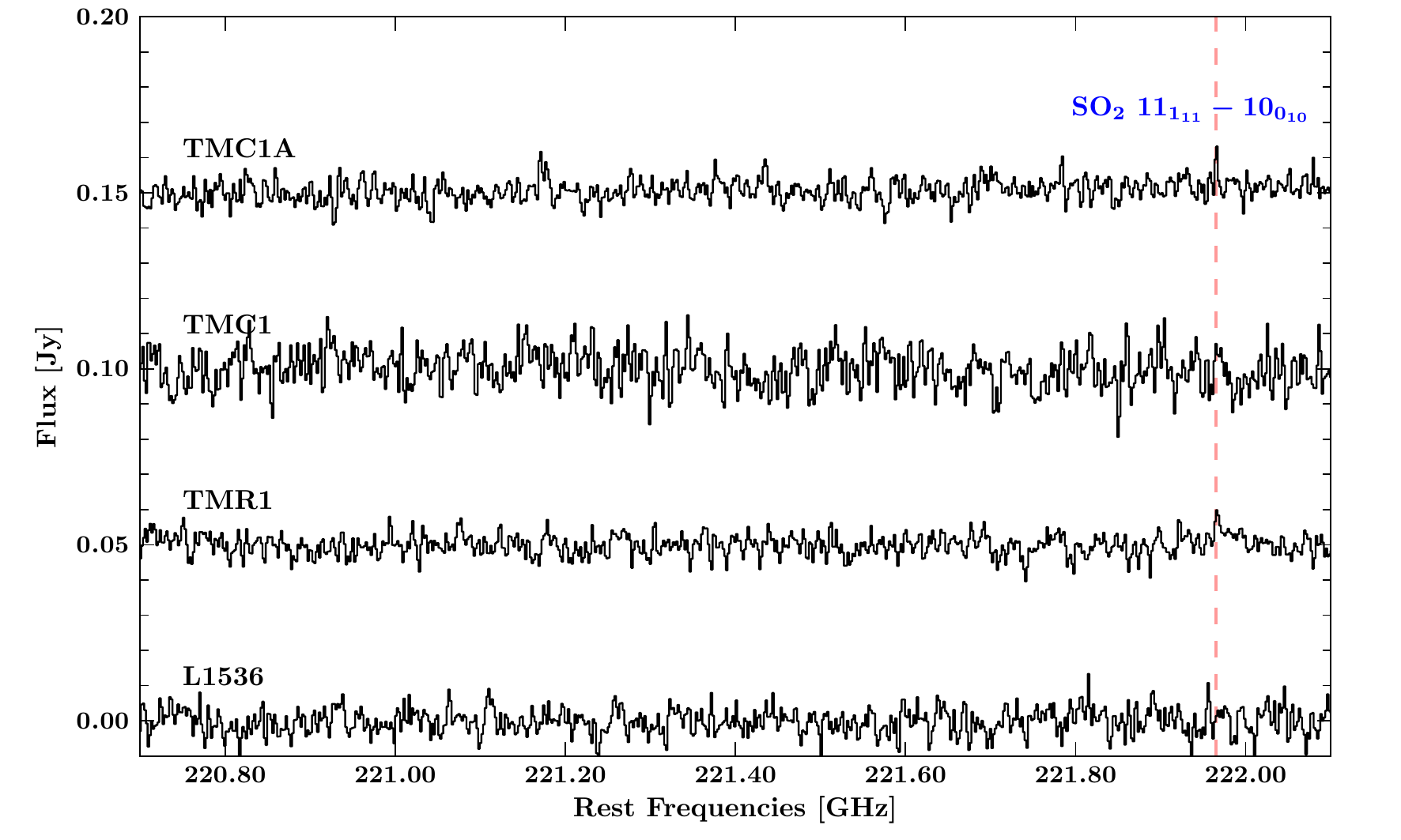}
 \caption{Integrated flux within a 1$\arcsec$ box around each source in the 
WideX spectra between 220.7--222.1 GHz.}
  \label{fig:allwidexspecs1}
\end{figure*}

\begin{figure*}[htbp]
 \centering
 \includegraphics{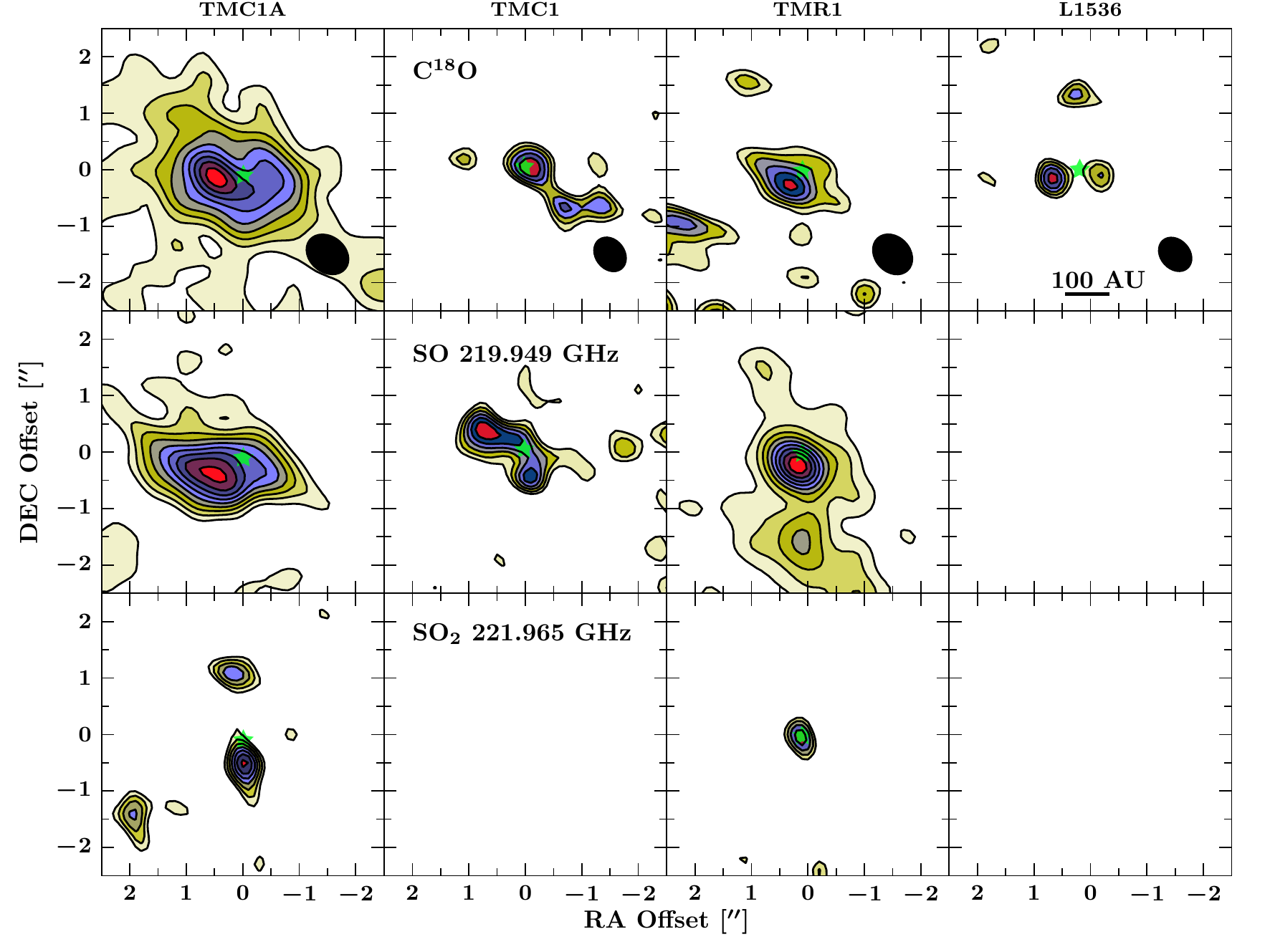}
 \caption{Moment 0 maps of \ceo\ $J=$2--1, SO $5_6 - 4_5$ and SO$_2$ $11_{1_11} 
- 10_{0_{10}}$ toward all sources from the WideX data.  The contours are 
drawn at 10\% of the peak starting at 3$\sigma$. Green star indicates the 
continuum position and black ellipses show the synthesized beams.}
  \label{fig:allwidexmaps}
\end{figure*}

Spectra toward all sources integrated within a 3$\arcsec$ box around their 
continuum positions are shown in Fig.~\ref{fig:allwidexspecs}.  The integrated 
flux density maps toward \ceo, SO $5_6 - 4_5$ and SO$_2$ $11_{1_11} - 
10_{0_{10}}$ (221.96521 GHz) are shown in Fig.~\ref{fig:allwidexmaps}.   
The SO emission toward all sources is much more extended compared to the \ceo\ 
emission, while the SO$_2$ emission is compact. 


\section{L1536}\label{app:l1536}

\begin{figure*}[htbp]
 \centering
 \includegraphics{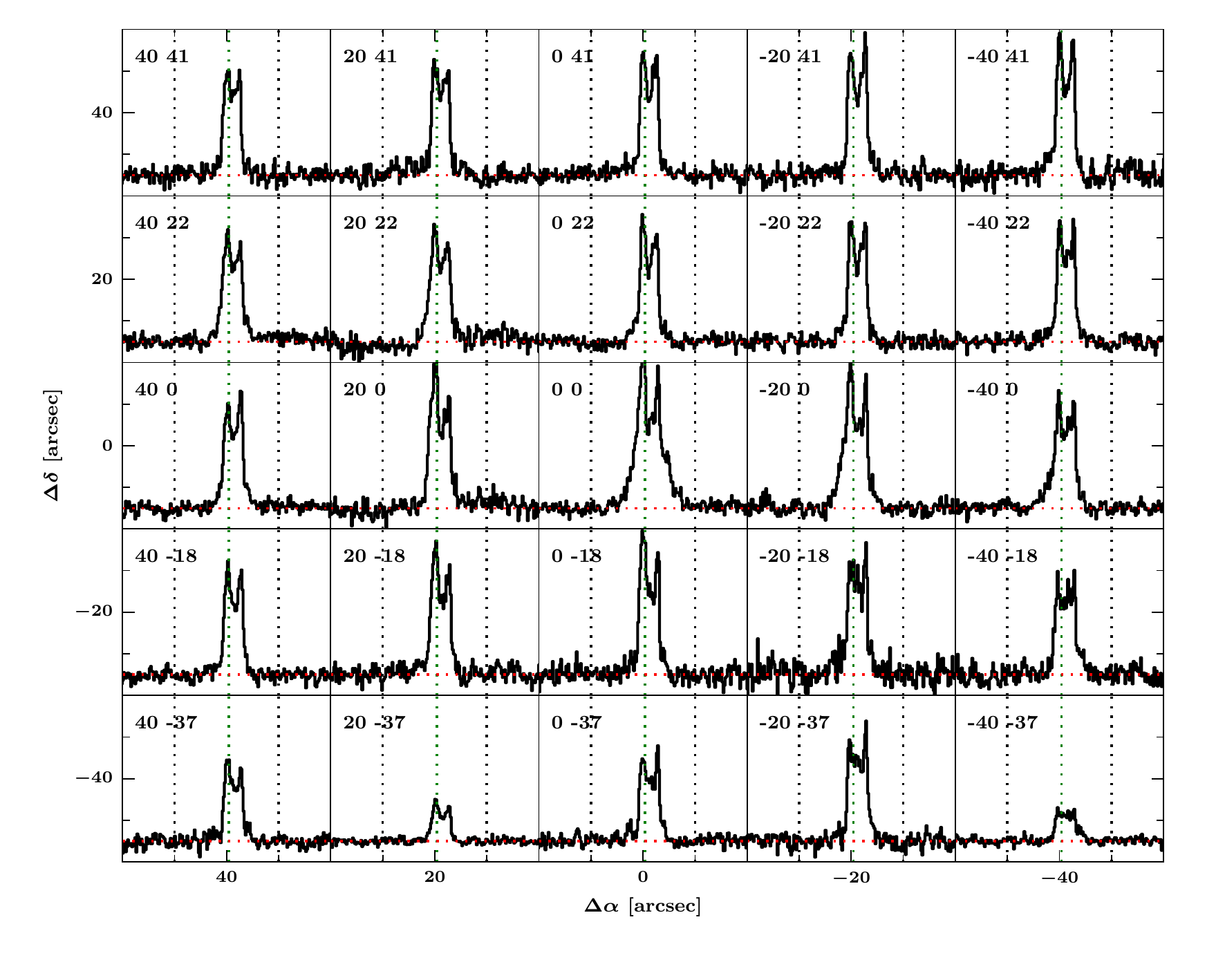}
 \caption{ \mco\ $J = $3--2 spectral map obtained with the JCMT from -5 \kms\ to 
15 \kms\ and intensities from -0.5 K to 3.5 K.  The vertical dotted lines show 
0, 5.2, and 10 \kms\ for guidance. }
\label{fig:l1536specmap}
\end{figure*}

Figure~\ref{fig:l1536specmap} shows the \mco\ $J=$3--2 spectral map obtained 
with JCMT, where there is no indication of classical bipolar outflows are 
present.


\end{document}